\documentclass[superscriptaddress,floatfix,nofootinbib,longbibliography,prx,twocolumn]{revtex4-1}
\usepackage[bookmarks=false]{hyperref}
\usepackage{graphicx,amsfonts,amssymb,amsmath,hypcap,enumerate}
\usepackage{slashed}
\usepackage{amsmath,amssymb,bm,graphicx,dsfont}
\usepackage[latin9]{inputenc}
\usepackage{amsmath}
\usepackage{slashed}
\usepackage{dsfont}
\usepackage{amstext}
\usepackage{amssymb}
\usepackage{amsbsy}
\usepackage{amsthm}
\usepackage{epsfig}
\usepackage{graphicx}
\usepackage{bbm}
\usepackage{hyperref}
\usepackage{color}
\usepackage{booktabs}
\usepackage{array}
\usepackage{slashed}
\newcommand{\be}{\begin{equation}}
\newcommand{\ba}{\begin{align}}
\newcommand{\ee}{\end{equation}}
\newcommand{\bea}{\begin{eqnarray}}
\newcommand{\eea}{\end{eqnarray}}
\newcommand{\beq}{\begin{equation}}
\newcommand{\eeq}{\end{equation}}
\newcommand{\beqn}{\begin{eqnarray}}
\newcommand{\eeqn}{\end{eqnarray}}

\newcommand{\ra}{\rangle}

\newcommand{\lp}{\left(}
\newcommand{\rp}{\right)}

\def\nn{\nonumber\\}

\begin{document}
\title{Correlated Fragile Topology: a Parton Approach}
\author{Katherine Latimer}
\affiliation{Perimeter Institute for Theoretical Physics, Waterloo, Ontario N2L 2Y5, Canada}
\affiliation{Department of Physics, University of California, Berkeley, California 94720, USA}
\author{Chong Wang}
\affiliation{Perimeter Institute for Theoretical Physics, Waterloo, Ontario N2L 2Y5, Canada}
\date{\today}

\begin{abstract}
{
\noindent
A fragile topological insulator (FTI) can be viewed as an almost-atomic insulator, with emergent negative charges localized at certain real space points, even though the underlying lattice Hilbert space contains only positively charged states. Fragile topology in free fermion systems has been fruitfully studied using modern topological band theory. However, the concept is well defined even for strongly correlated systems, and fragile states that cannot be realized within free fermion band theory exist abundantly. In this work we propose a rather general parton construction for such correlation-enabled FTIs. In our parton construction the associated gauge symmetries are completely Higgsed, resulting in only short-range entangled states. The effective negative charges in the FTIs emerge naturally as the remnants of negatively charged partons. For spinful electrons with $SU(2)$ spin-rotation symmetry, the fragile phases can be viewed as FTIs of charge-$2$, spin singlet bosonic Cooper pairs. 
We discuss a few examples of correlated FTIs for both spinless and spinful fermions, including some ``featureless Mott insulators" on the honeycomb lattice previously discussed in the literature.
 }
\end{abstract}

\maketitle


\section{Introduction}

The interplay between symmetry and entanglement plays a major role in modern many-body quantum physics. An elegant illustration of such interplay comes from the study of symmetry-protected topological (SPT) phases of matter\cite{ChenCohomology}. By definition, SPT phases cannot be adiabatically deformed into unentangled tensor product states (or ``atomic insulators") as long as certain symmetries are unbroken. Prototypical examples include the Haldane-AKLT spin chain and the free fermion topological insulators. 

\begin{center}
\begin{figure}[ht]
{
\includegraphics[width=1\columnwidth]{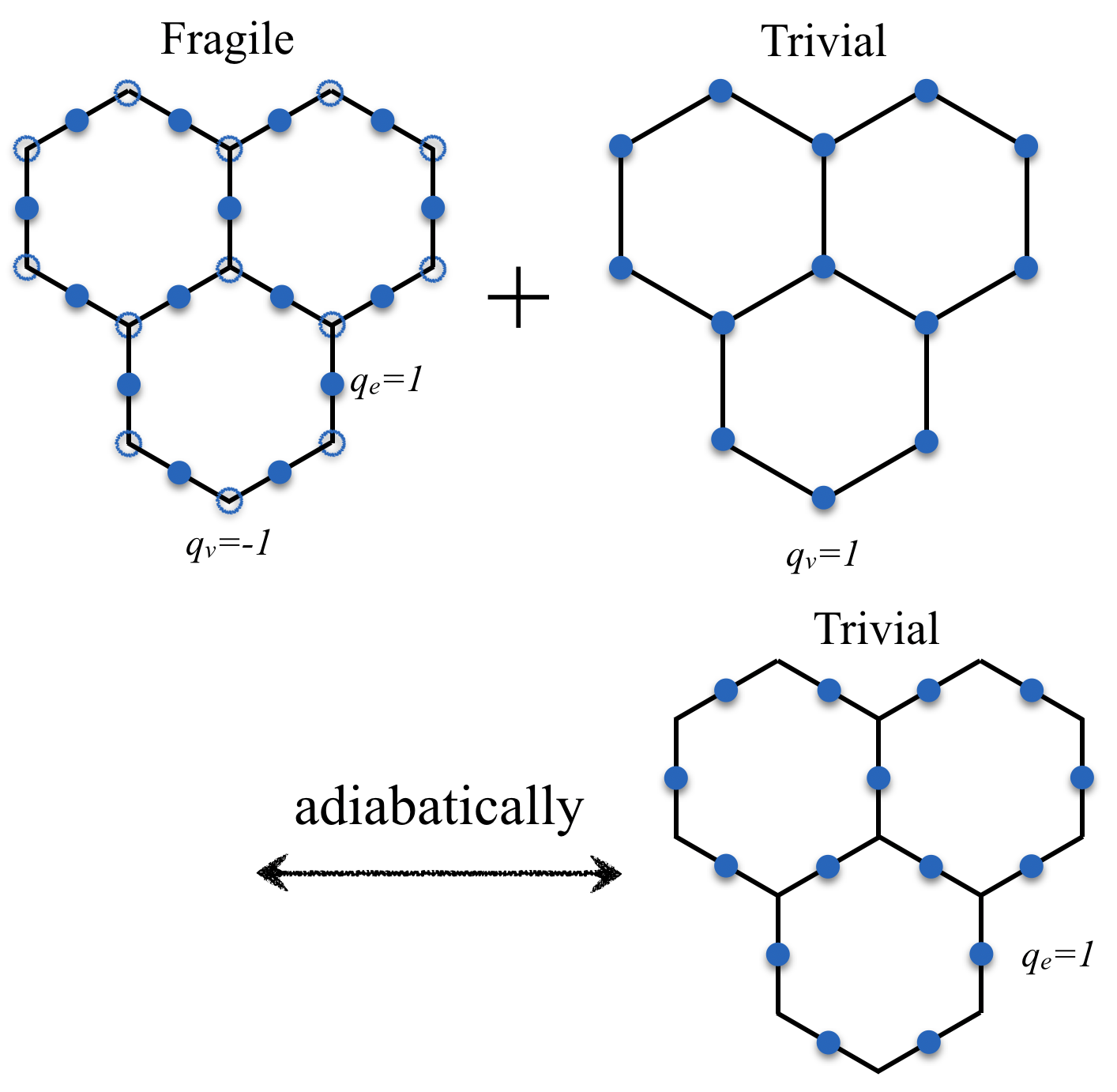}}
\label{FTIillu}

\caption{
An example of fragile topology on honeycomb lattice with spinless electrons at half-filling, i.e. $\langle c^{\dagger}_ic_i\rangle=1/2$ for every lattice site where $c_i, c_i^{\dagger}$ are the electron annihilation and creation operators on site $i$. In the fragile phase there is a positive charge $q_e=1$ localized on every edge of the lattice, and an emergent negative charge $q_v=-1$ localized on every vertex. The emergent negative charge is a result of many-body entanglement. If we ``neutralize" the negative charge by coupling the fragile phase with another trivial (i.e. unentangled) atomic insulator with only $q_v=1$, the total system can be adiabatically deformed to an atomic limit with positive charge only ($q_e=1$). As we will see in Sec.~\ref{honeycomb}, this fragile phase can be realized only with strong electron-electron interactions. In fact its trial wavefunction has been constructed in Ref.~\cite{RanFeatureless} as an example of ``featureless Mott insulator".
}
\end{figure}
\end{center}

Motivated by the study of electronic insulators, much attention has been focused on systems with a global symmetry group $G$ that contains a $U(1)$ subgroup (i.e. charge conservation). A novel subclass of SPT phases in such systems was proposed recently, known as \emph{fragile} topological insulators (FTIs)\cite{Pofragile, Cano_2018}. An FTI is nontrivial on its own, but can be adiabatically and symmetrically deformed to an atomic insulator when coupled with another atomic insulator. 
An atomic insulator is labeled by the content of localized particles on each real space point within a unit cell, i.e. how many particles (charges) are localized at each point and what quantum numbers (such as lattice angular momentum) these particles carry. The charges are typically localized at some high-symmetry points in real space, also known as Wyckoff positions. A truly unentangled atomic insulator in a system of positively charge particles would contain a non-negative electric charge at each Wyckoff position. In contrast, an FTI has both positively and negatively charged Wyckoff positions, and can be deformed to a truly unentangled atomic state only when it is coupled to another positively charged atomic insulator that ``neutralizes" all the negative charges of the FTI. The emergence of negative charges in an FTI is a result of the nontrivial (symmetry-protected) many-body entanglement. In Fig.~\ref{FTIillu} we illustrate these concepts with an example of spinless fermions on a honeycomb lattice at half-filling -- this example will play an important role in the rest of this paper. 

Fragile topology was first proposed within the free fermion band theory\cite{Pofragile}, and has played a surprisingly interesting role in recent study of the magic angle twisted bilayer graphene\cite{TBGFragile}. The concept of fragile topology based on adiabatic deformations, however, is well defined even for strongly correlated systems and has been discussed in detail in Ref.~\cite{ElsePoWatanabe} (see also Ref.~\cite{shiftinsulator}). Interestingly, there are strongly correlated FTIs that cannot be realized within free fermion band theory, as first demonstrated also in Ref.~\cite{ElsePoWatanabe}. In fact, it is well known that the example illustrated in Fig.~\ref{FTIillu} on honeycomb lattice cannot be realized within band theory. The study of such correlated FTIs requires tools beyond band theory. In particular, it is desirable to have a better understanding of the emergence of the localized effective negative charges. Ref.~\cite{ElsePoWatanabe} approached this problem using the trick of charge conjugation operation, in which negative charges appear naturally as the conjugate of positive ones. 


In this work we develop a complementary approach to construct and analyze correlated FTIs using the parton trick (for reviews of the trick see Ref.~\cite{wenbook, LeeNagaosaWen}). Roughly speaking, partons are fictitious particles introduced to capture the non-trivial many-body entanglement in exotic phases. For example, as we will show in detail in Sec.~\ref{honeycomb}, for the FTI in Fig.~\ref{FTIillu} the relevant parton construction starts by re-writing the electron anihilation operator as
\be
\label{Parton1}
c=\psi_1\psi_2\psi_3,
\ee
where $\psi_{1,2,3}$ are fermionic partons. Many interesting phases can be constructed out of simple mean-field states of the partons -- the most familiar examples are fractional quantum Hall states and quantum spin liquids. Depending on the details of the mean field states, the $\psi_{1,2,3}$ partons can be assigned different electric charges, satisfying the constraint $q_1+q_2+q_3=1$ (we define electron charge to be $1$). The charge assignment corresponding to the FTI in Fig.~\ref{FTIillu} is $q_1=q_2=1$ and $q_3=-1$ -- in this case $\psi_3$ is an emergent particle with negative charge, and is ultimately responsible for the effective negative charges at the vertices.  More generally, in our parton approach the emergent negative charges in FTIs can be simply attributed to some negatively charged partons. This provides a relatively simple picture of fragile topology in correlated systems. Amusingly, this state has been discussed in Ref.~\cite{RanFeatureless} as an example of a ``featureless Mott insulator"\cite{KimchiFeatureless,JianFeatureless} (defined as a fully symmetric and gapped state without intrinsic topological order) on a half-filled honeycomb lattice, before the notion of fragile topology even appeared. 
Armed with this understanding, we can easily construct other fragile phases with different charge distribution patterns on various lattices. We discuss several illuminating examples in Sec.~\ref{examples}. As we will see, a unifying feature of these parton constructions is that the gauge symmetries associated with the partons are completely broken (or Higgsed), so that the resulting states are all short-range entangled, with no active gauge field remain at low energy. 

In Sec.~\ref{spinhalf} we extend our discussion to spin-$1/2$ electrons with $SU(2)$ spin rotation symmetry. The only difference is that localized charges must form spin singlets to preserved the $SU(2)$ symmetry, which in turn requires the charge at each point to be an even integer. In other words, atomic and fragile insulators of spin-$1/2$ fermions can all be viewed as those formed by tightly bound charge-$2$, spin singlet bosonic Cooper pairs. Interestingly, as we will show, some of these fragile Cooper pair insulators can also be adiabatically deformed to a ``strong Mott insulator", defined as a charge insulator with exactly one localized electron per lattice site (the original definition of a Mott insulator). A strong Mott insulator essentially corresponds to a pure spin model, although the Hamiltonian can be more complicated than the simple Heisenberg model. In this limit, there is no effective negative charge: instead, the quantum entanglement is stored purely in the spin degrees of freedom, and such insulators can be viewed as quantum paramagnets. We dub such fragile insulators ``fragile topological paramagnets". We will also show that some of the ``featureless quantum paramagnets" (again defined as fully symmetric and gapped states without intrinsic topological order) on honeycomb lattice discussed in Ref.~\cite{RanFeatureless} are in fact early examples of fragile topological paramagnets. In fact, we will show that any featureless quantum paramagnets realized on a half-filled honeycomb lattice, with one orbital per site, is necessarily fragile.

Before starting our detailed discussions, we first review several basic notions required to precisely define fragile topology in correlated systems in Sec.~\ref{BasicNotions}, in a language that will be convenient for our purpose.

\section{Basic Notions}
\label{BasicNotions}

The goal of this section is to review the precise definition of fragile topology in correlated systems, in a language convenient for our later discussions. 

First, we will consider systems with a global $U(1)$ charge symmetry, together with other lattice and internal symmetries. We assume the Hilbert space to be local, in the sense that it is a tensor product of local Hilbert spaces.  Crucially, the Hilbert space should contain only states with positive (or at least zero) $U(1)$ charge. In other words, the Hilbert space consists of a vacuum with non-negative charge, and positively charged particles on top of the vacuum. We dub this property of Hilbert spaces \textit{charge positivity}. If we have a lattice system with finite Hilbert space dimension per unit cell, then we can always make it charge positive by redefining the vacuum charge. For example, a honeycomb lattice with one electron orbit defined on each vertex is positively charged if we define the empty state (particle vacuum) to have zero charge -- this is the Hilbert space we will be using to study the honeycomb example in Sec.~\ref{honeycomb}. In contrast, if we have a quantum rotor on each vertex, defined using a phase variable $e^{i\theta}$ and its conjugate angular momentum $L=-i\partial_{\theta}$, the Hilbert space is not positively charged since the charge $L$ can take any integer value.

Next, we will consider ground states that are (a) short-range entangled and (b) do not have protected gapless (or more precisely anamalous) boundary modes. As argued in Ref.~\cite{song_2017}, such states can be ``dimensionally reduced" to a stack of $(0+1)d$ states. To describe such a state, it is sufficient to enumerate the symmetry quantum numbers associated with each Wyckoff position in real space. For example, consider a Wyckoff position $i$ invariant under a $C_{n_i}$ point group symmetry in $2D$. We can associate to it a $U(1)$ charge $q_i$ and a $C_{n_i}$ eigenvalue (angular momentum) $l_i$ (and other internal symmetry quantum numbers if necessary). The interpretation is that if we keep only the internal symmetries and $C_{n_i}$, we can completely dis-entangle the ground state, and the remaining product state will have total $C_{n_i}$ eigenvalue $l_i$ and total charge $q_i$ (mod $n_i$). Alternatively, one can also define the effective charge via topological response: the effective charge can be determined from either the fractional charge carried by a lattice disinclination\cite{shiftinsulator} or the lattice angular momentum carried by a magnetic flux quanta\cite{shiftinsulator,Songetal}. In either approach, it is clear that the effective charge $q_i$ is defined only mod $n_i$. Similar logic also applies in $3D$: the charge $q_i$ at point $i$ is defined mod $n_i$ where $n_i$ is determined by the point group symmetry at $i$. In this paper we will mostly focus on $2D$ systems, except in Sec.~\ref{3Dexample}.

Although each $q_i$ is defined only mod $n_i$, the total charge in each unit cell is given by the filling factor $\nu\in\mathbb{N}$:
\be
\label{totalcharge}
\sum_{i\in \textit{unit cell}}q_i=\nu.
\ee

A strongly correlated insulator is defined to be fragile if 
the solution of the above equation necessarily contains some negative $q_i$. For example, take the state described in Fig.~\ref{FTIillu}: on each honeycomb vertex we have $q_v=-1$ (mod $3$), on each edge we have $q_e=1$ (mod $2$) and on each hexagon center we have $q_h=0$ (mod $6$) -- recall that the ambiguity $n$ is determined by the rotation symmetry $C_n$ at each Wyckoff position. The total charge filling per unit cell is $2q_v+3q_e+q_h=1$. It is then easy to see that at least one of these charge numbers has to be negative.

Since our Hilbert space is positively charged, any un-entangled tensor product state can only host a non-negative charge at each Wyckoff position. Therefore the existence of an effectively negative charge at certain Wyckoff position implies that the state cannot be completely trivial -- some remnant quantum entanglement has to be responsible for the emergence of such negative charge. This is the significance of fragile topology in generic interacting systems.

In summary, an insulator in a generic interacting system is defined to be fragile if
\begin{enumerate}
\item the underlying Hilbert space is positively charged,

\item the ground state is short-range entangled and has no protected gapless (or anomalous) boundary mode. This implies that $q_i$ is well defined for each Wyckoff position $i$.

\item Eq.~\eqref{totalcharge} has no non-negative solution, i.e. the effective charges at some Wyckoff positions are necessarily negative.
\end{enumerate}

A consequence of this definition is that many phases that are fragile within free fermion theory become non-fragile once strong interactions are introduced\cite{shiftinsulator}. This is because while the angular momentum carried by each individual localized particle is well defined within free fermion theory, only the total angular momentum at each point remains well-defined with interactions.

Here we are interested in FTIs that cannot be realized within free fermion band theory. In principle there are two slightly different notions of such correlated FTIs. An FTI may be realizable within band theory if the Hilbert space of each unit cell is sufficiently large, but requires strong correlation when the lattice Hilbert space is more restricted. A stronger type of correlated FTI would require strong interactions for any lattice Hilbert space that is consistent with the symmetries and charge positivity -- we call such FTIs \textit{intrinsically correlated}. Next in Sec.~\ref{honeycomb} we will discuss an example of intrinsically correlated FTI. 

\section{An example on honeycomb lattice}
\label{honeycomb}

We now discuss the particular example of the FTI shown in Fig.~\ref{FTIillu}. It is well known that this insulator (or any other insulating state) cannot be realized in a free fermion tight-binding model defined on a honeycomb lattice with all the standard symmetries ($U(1)$ charge conservation, lattice translations and rotations and time-reversal). It is also true that even if we extend the Hilbert space, for example by introducing additional orbits at the edge or plaquette centers, this particular FTI is still unrealizable with free fermion models. This can be shown using the tool of symmetry indicators (also known as topological quantum chemistry)\cite{indicators,Bradlyn2017,Slager17,Slager19}. To make our discussion self-contained, we briefly review the symmetry indicator argument in Appendix~\ref{Sindicator}.\footnote{This makes our FTI more intrinsically correlated than the simplest ``featureless Mott insulator"\cite{KimchiFeatureless}, which has one positive charge at each hexagon plaquette center and no negative charge (and is therefore not fragile). Although both states require strong interactions on a half-filled honeycomb lattice, the non-fragile state can be trivially realized within free fermion theory if we add an extra orbital at each hexagon center.} 

We now proceed to construct the FTI state using the parton trick. We shall work with the simple Hilbert space with one fermion orbit $c_i$ defined on each lattice vertex $i$. We start from the parton decomposition in Eq.~\eqref{Parton1}. However we will use a slightly different notation for later convenience:
\be
\label{Parton2}
c=f_0^{\dagger}f_+f_-,
\ee
where $f_{0,+,-}$ are canonical fermion operators. 
Eq.~\eqref{Parton2} differs from Eq.~\eqref{Parton1} by charge-conjugating one of the partons. The physical states are represented in the parton Hilbert space as
\bea
\label{physicalH}
|0\ra&=&f^{\dagger}_0|\Omega\ra, \nn
c^{\dagger}|0\ra&=&f_-^{\dagger}f_+^{\dagger}|\Omega\ra,
\eea
where $|\Omega\ra$ is the parton vacuum state: $f_{0,+,-}|\Omega\ra=0$. 

We further define another set of fermions $f_{1,2,3}$:
\be
\label{orbitals}
f_{a}=\frac{1}{\sqrt{3}}\Big(f_0+e^{-ia2\pi/3}f_++e^{ia2\pi/3}f_-\Big),
\ee
where $a\in \{1,2,3\}$.

\begin{figure}[h]
  \centering
    \includegraphics[width=.5\textwidth]{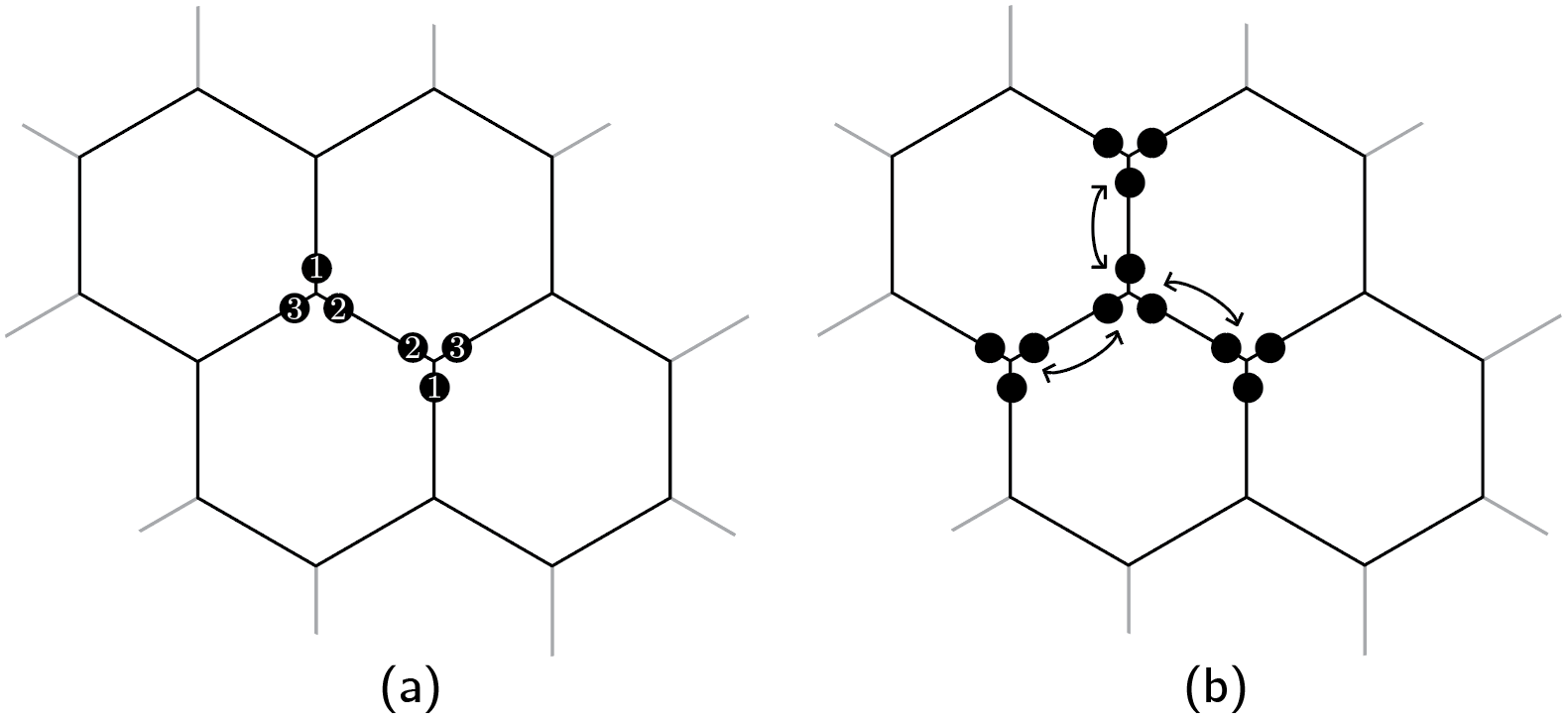}
    \caption{(a) Parton orbitals for Example 1. (b) Ground state of mean-field Hamiltonian (\ref{hmf}).}
  \label{hclatt1}
\end{figure}

Our parton mean field Hamiltonian takes the form
\begin{equation}
    \label{hmf}
    H_{MF}=\sum\limits_{ij}\sum\limits_{ab}(t_{ij})_{ab}f^{\dagger}_{a,i}f_{b,j}+h.c.,
\end{equation}
where $i$ and $j$ are site labels and $(a,b)\in 1,2,3$ label the parton orbitals at each site as defined in Eq.~\eqref{orbitals}. The hopping parameters $t_{ij}$ (determined by the form of the mean field) are given by:
\begin{align}
    \label{ansatz}
    t_{ij}=\begin{pmatrix}
        -t&0&0\\
        0&0&0\\
        0&0&0\\
    \end{pmatrix}&\quad ij=\includegraphics[scale=0.3, trim=0 1cm 0 0]{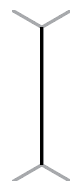}\nonumber\\
    \begin{pmatrix}
        0&0&0\\
        0&-t&0\\
        0&0&0\\
    \end{pmatrix}&\quad ij=\includegraphics[scale=0.3, trim=0 0.8cm 0 0]{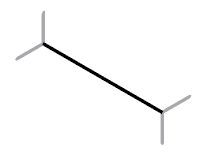}\nonumber\\
    \begin{pmatrix}
        0&0&0\\
        0&0&0\\
        0&0&-t\\
    \end{pmatrix}&\quad ij=\includegraphics[scale=0.3, trim=0 0.8cm 0 0]{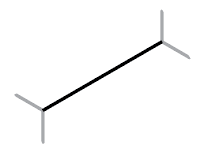},\nonumber\\
\end{align}
where $t$ is a positive constant. The interpretation is that each orbital $f_a$ only hops along a specific bond as in Fig.~\ref{hclatt1}. By Eq.~\eqref{orbitals} we can interpret $f_{0,+,-}$ as orbitals with lattice angular momenta (eigenvalues under site-centered $C_3$ rotations) $\{1,e^{i2\pi/3},e^{-i2\pi/3}\}$.

The mean field Hamiltonian Eq.~\eqref{hmf} is clearly invariant under lattice and time-reversal operations -- the only minor subtlety is that mirror reflections exchange $f_+$ and $f_-$. Therefore, in order for $c$ to be invariant, all the $f$ fermions in Eq.~\eqref{Parton2} should pick up an extra minus sign under mirror reflections. It also has the charge $U(1)$ symmetry, with $f_{1,2,3}$ all assigned with charge $q=1$ -- this requires $f_{0,+,-}$ to also have charge $q=1$, fully consistent with the parton decomposition Eq.~\eqref{Parton2}. 

The ground state of $H_{MF}$ is a tensor product of localized hopping partons, as visualized in Fig.~\ref{hclatt1}(b). Such a state corresponds to very localized interactions between the particles; in principle, one could allow the partons more freedom to disperse throughout the lattice at the zeroth-order mean-field level, but the qualitative picture (namely the topological aspects) would not change.

We now analyze the state beyond mean-field level. There are two complementary ways to do so. One can take the mean field ground state and project back to the physical Hilbert space through Eq.~\eqref{physicalH}. The resulting state is in general complicated but can be analyzed numerically. For the particular state from the mean field Eq.~\eqref{ansatz} the numerical analysis has been done in Ref.~\cite{RanFeatureless}, and it was found that this parton state is short-range entangled (namely has no topological or gapless order) and has no spontaneously broken symmetry. Alternatively, one can analytically include the fluctuation beyond mean field by examining the gauge symmetries associated with the mean field state.  

Gauge symmetry can be analyzed as follows: the parton decomposition Eq.~\eqref{Parton2} comes with an $SU(3)$ gauge symmetry
\be
\label{newsu3}
\lp\begin{array}{c}
   f_0^{\dagger} \\
   f_+ \\
   f_- \end{array}\rp \to U\lp\begin{array}{c}
   f_0^{\dagger} \\
   f_+ \\
   f_- \end{array}\rp,
\ee
where $U$ is an $SU(3)$ matrix. An important question is: how much of this gauge group is kept unbroken by the mean field Hamiltonian Eq.~\eqref{hmf}? The gauge symmetry preserved by the mean field state is also known as the invariant gauge group (IGG)\cite{wenbook} or low-energy gauge group\cite{LeeNagaosaWen}. In general, at low energy and long distance, a fluctuating gauge field with the gauge group IGG will couple to the partons and may significantly modify the mean field dynamics. However, by inspecting Eq.~\eqref{hmf} and \eqref{ansatz}, there is no nontrivial $SU(3)$ gauge transform that leaves the mean field invariant -- we show this in detail in Appendix.~\ref{IGG}. 

Since the IGG is completely trivial, there is no fluctuating gauge field needed in the low energy description. The physical state is therefore gapped and fully symmetric just like the mean field parton state. The triviality of the gauge group also means that there is no intrinsic topological order -- the state is short-range entangled. The trivial IGG also implies that the symmetry transformation properties of the $f$ fermions determined above (like their $C_3$ eigenvalues) are unambiguous, since there is no gauge symmetry left to redefine the symmetry transformations.

The parton mean field theory does not lead to any protected gapless edge mode. The triviality of IGG implies that the physical state also has no protected gapless edge mode. We therefore expect the resulting insulator to be either trivially atomic or at most fragile. Since gauge fluctuations are suppressed, the effective electric charge at each point in real space can be determined directly from the parton mean field theory. As illustrated in Fig.~\ref{hclatt1}, effectively there is one $f$ fermion localized on each bond of the honeycomb lattice, so we assign each bond center an electric charge $q_e=1$. The sign of $t$ in Eq.~\eqref{ansatz} is such that the $f$ fermion living on the bond has inversion eigenvalue $1$. On each lattice site effectively there is no $f$ fermion remained, so the effective charge on each site is the same as the parton vacuum $|\Omega\rangle$ -- but what is this parton vacuum charge? By Eq.~\eqref{physicalH} the parton vaccum $|\Omega\ra$ should have charge $q(|\Omega\ra)=q(|0\rangle)-1$, where $q(|0\rangle)$ is the charge of the physical vacuum. Now we define $q(|0\rangle)=0$ so that the Hilbert space is potiviely charge (as discussed in Sec.~\ref{BasicNotions}). The effective charge at each lattice vertex is therefore $q_v=-1$. The total charge in each unit cell is $3-2=1$, in agreement with the microscopic charge filling. As we promised earlier, this is exactly the state illustrated in Fig.~\ref{FTIillu}. The effective negative charge simply comes from the negatively charge parton vacuum $|\Omega\ra$, or equivalently a negatively charge parton $f_0^{\dagger}$ in Eq.~\eqref{Parton2}.

\section{More examples}
\label{examples}
Below, we generalize our construction used in Sec.~\ref{honeycomb} and present further examples of correlated FTIs, in spinless fermion lattice systems with fillings that do no allow band insulators (so that the resulting insulators constructed clearly require strong correlations).\footnote{One could also go through a similar exercise as in Appendix~\ref{Sindicator} to check if the FTIs in this section are intrinsically correlated (independent of Hilbert space structures). However the task becomes quite tedious and un-illuminating for more complicated examples so we do not attempt to do it here.}  

Let us first outline the general strategy we will be employing, systematizing what was done in Sec.~\ref{honeycomb}. Our general approach is to draw a distribution of positive and negative charges in the unit cell that gives the desired net filling. We interpret those charges as partons constitutive of the physical fermion field and rearrange them within the lattice, making sure to retain correct quantum numbers for the physical particles.

 In all cases, we use a parton decomposition with $2m+1$ fermion partons
\be
c=f_1f_2...f_{m+1}f^{\dagger}_{m+2}f^{\dagger}_{m+3}...f^{\dagger}_{2m+1},
\ee
with each fermion parton $f$ carrying electric charge $1$. This introduces an $SU(2m+1)$ gauge symmetry. We then spread the $f$ fermions to form localized positive charges at various points within each unit cell. On a Wyckoff position $i$ where the physical $c$ fermion is defined, however, the effective charge is $n_i=n_f-m$ since the physical charge vacuum state has the fermions $\{f_{m+2},...f_{2m+1}\}$ filled. This on-site effective charge can then be negative with appropriate choices of $n_f$. In this construction the effective negative charges only sit on the Wyckoff positions where the physical fermions are defined. This does not loose any generality since we can always adiabatically move all the negative charges to those positions, possibly by creating more positive charges elsewhere\cite{LatticeHomotopy}. Finally, if there is still a nontrivial gauge symmetry (IGG) that is not broken, we add various local symmetry-preserving terms to the mean field Hamiltonian to ensure that the $SU(2m+1)$ gauge symmetry is completely Higgsed -- the Higgs terms will not affect the charge distribution as long as they are not large enough to close the mean field gap. Typically the Higgs terms are not completely on-site -- for example, in Sec.~\ref{honeycomb} they live on the links. Intuitively such non-on-site Higgs terms are the sources of the remaining short-range entanglement in the fragile topological phases.

\subsection{Honeycomb lattice}
With our first example on the honeycomb lattice in Sec.~\ref{honeycomb}, we have mostly worked in the real space orbital basis, explicitly transforming to the angular momentum basis for clarity of some expressions. However, it will be most convenient for the host of proceeding examples to work in the basis of irreducible representations (irreps) of the relevant site symmetry (i.e. point) groups on the various lattices, since this allows one to see immediately how both the partons and the physical particles transform under symmetry operations.

\subsubsection{A simple example}
\label{simpleEx}

\begin{figure}[h]
  \centering
    \includegraphics[width=.5\textwidth]{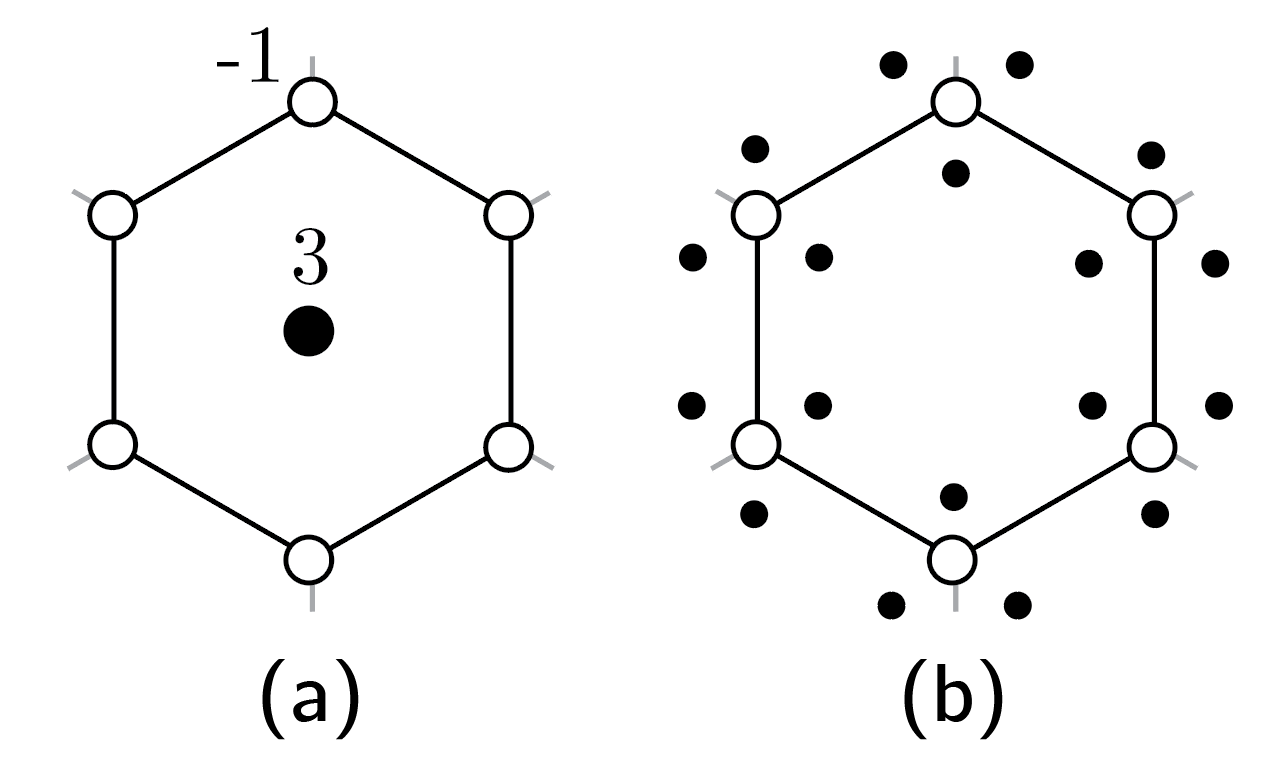}
    \caption{(a) Physical $U(1)$ charge distribution for second parton construction on the honeycomb lattice. (b) Distributing parton orbitals in cell centers to orbitals centered around the sites. Each site has three nearest cell center neighbors.}
  \label{hclatt2}
\end{figure}

We can see how this works for another simple state on the honeycomb lattice, with a $U(1)$ charge distribution shown in Fig.~\ref{hclatt2}(a), with charge $3$ at each plaquette center and $-1$ on each site. To motivate the parton construction, we imagine ``smearing" three fermionic partons from each plaquette center into six discrete orbitals within the plaquette associated with the hexagonal vertices, as shown in Fig.~\ref{hclatt2}(b). The three orbitals at each lattice site decompose into irreps of that site symmetry group, $C_{3v}$, as:
\begin{equation}
    A_1\oplus E
\end{equation}
For this and all ensuing examples, representations labeled by $A$ and $B$ are one-dimensional, and those labeled by $E$ are two-dimensional, and denote the independent components of the two-dimensional $E_1$ representation as $E_1^+$ and $E_1^-$. This motivates the same parton decomposition as in our first example, now written in the new notation as:
\begin{equation}
    c_s = f^{\dagger}_{s, A_1}f_{s, E^+}f_{s, E^-},
\end{equation}
where $s$ denotes the lattice site and we make the identifications $0\leftrightarrow A_1$, $+\leftrightarrow E^+$, and $-\leftrightarrow E^-$. 

These six orbitals enclosed within each hexagon in Fig~\ref{hclatt2}(b) decompose into irreps of the plaquette center's site symmetry group, $C_{6v}$, as:
\begin{equation}
    A_1\oplus B_1\oplus E_1\oplus E_2.
\end{equation}
Essentially they correspond to the six different $C_6$ angular momenta. A mean-field Hamiltonian yielding a symmetric ground state with a filling factor of three partons per unit cell is:
\bea
    H_{MF} = && -t\Bigg[\sum\limits_p\Big(f^{\dagger}_{p, A_1}f_{p, A_1} + f^{\dagger}_{p, E_1^+}f_{p, E_1^+} + f^{\dagger}_{p, E_1^-}f_{p, E_1^-}\Big)\Bigg] \nn && + \frac{t}{2}\sum_sf^{\dagger}_sf_s,
\eea
where $p$ labels a plaquette in the lattice, $s$ labels a site, and the final on-site sum ensures that the Fermi level lies within the energy gap. The fermions $f_{p, R}$ are linear combinations of the six real-space orbitals and transform like the irrep $R$ of $C_{6v}$; for example, $f_{p, A_1} = \frac{1}{\sqrt{6}}\Big(f_1 + f_2 + f_3 + f_4 + f_5 + f_6\Big)$. As in the previous case, the parton vacuum has electric charge $-1$. This leads to the charge distribution as in Fig.~\ref{hclatt2}(a).

\subsubsection{A slightly less simple example}
\label{lesssimpleEx}

\begin{figure}[h]
  \centering
    \includegraphics[width=.5\textwidth]{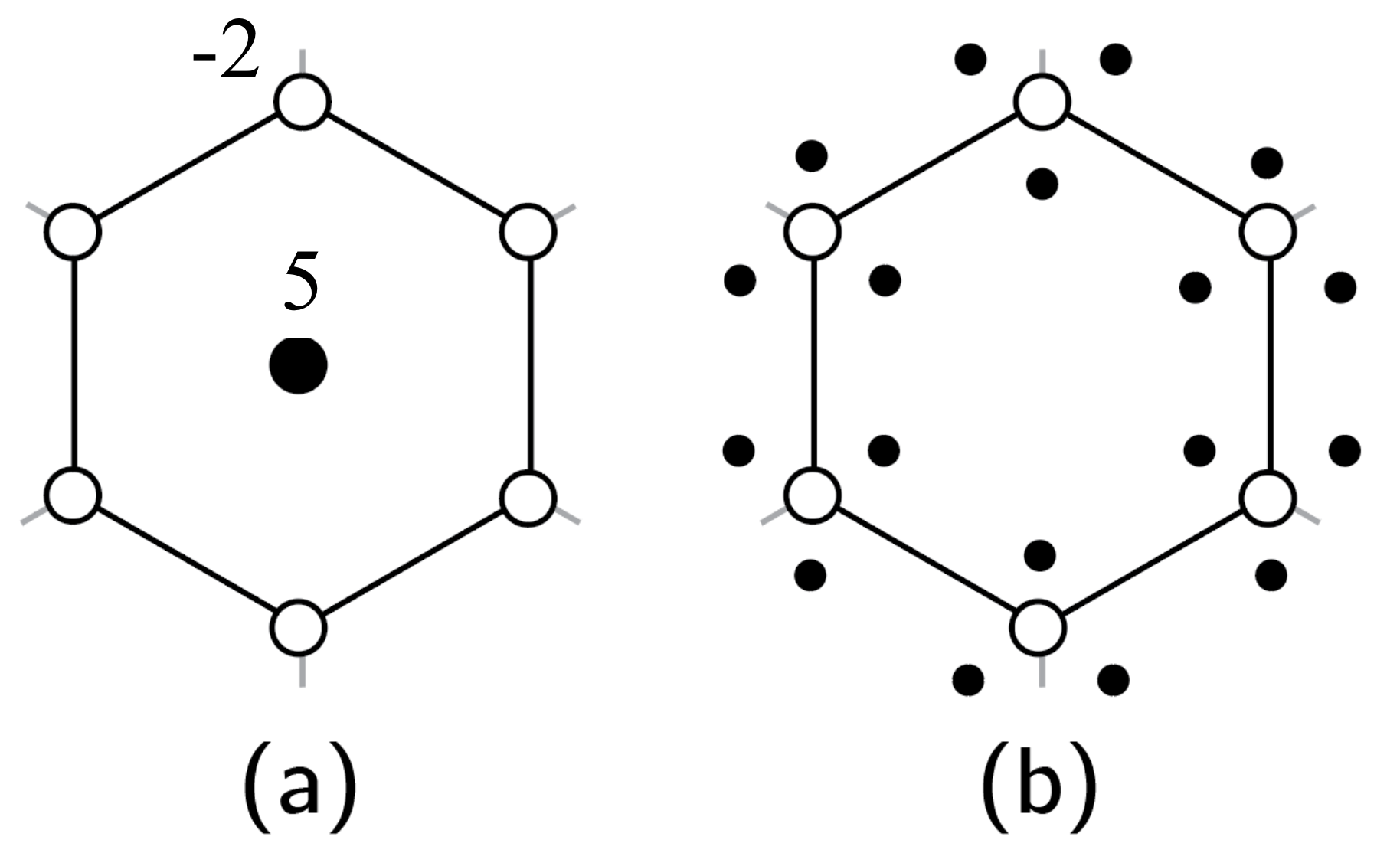}
    \caption{(a) Physical $U(1)$ charge distribution for third parton construction on the honeycomb lattice. (b) Distributing parton orbitals in cell centers to orbitals centered around the sites. Each site has three nearest cell center neighbors.}
  \label{hclatt4}
\end{figure}

We will examine one more example on the honeycomb lattice which introduces a new subtlety. The physical $U(1)$ charge distribution of the state is given in Fig.~\ref{hclatt4}(a), and we spread fermionic partons from the plaquette center in the same manner as above. However, the previous parton decomposition Eq.~\eqref{Parton2} will not be able to easily describe the $-2$ charge on the lattice sites since the parton vacuum only has $-1$ charge. We can overcome this by simply introducing two more partons, transforming under $C_{3v}$ group elements as the $E$ irrep, which we will label as the doublet $\Big(\psi_{s, E^+}, \psi_{s, E^-}\Big)$. A parton construction for a totally-symmetric physical fermion is:
\be
\label{Parton3}
c_s=f_{s,A_1}f_{s,E^+}f_{s,E^-}\psi^{\dagger}_{s,E^+}\psi^{\dagger}_{s,E^-}
\ee

Now to reproduce the charge pattern in Fig.~\ref{hclatt4}, we can put the $f$ fermions into filled orbits on the hexagons just as in the last example. The only difference is that we should fill five of the orbits, for example we can fill all the orbits except $f_{p,A_1}$. We then leave all the $\psi$ fermions un-occupied. Since the parton vacuum gives charge $-2$ on the lattice sites, this reproduces the desired charge distribution. However, this does not completely Higgs the gauge symmetry -- a $U(2)$ gauge symmetry on $(\psi_{E^+},\psi_{E^-})$ is not broken. Following the general strategy outlined at the beginning of this section, we should add terms that ``hybridize" the $\Big(f_{s, E^+}, f_{s, E^-}\Big)$ orbitals with the $\Big(\psi_{s, E^+}, \psi_{s, E^-}\Big)$ orbitals, in order to completely Higgs the gauge symmetry. This line of reasoning leads to the following mean field Hamiltonian:
\begin{align}
\label{hc4MF}
    H_{MF} = -t\Bigg[&\sum\limits_p\Big(f^{\dagger}_{p, A_2}f_{p, A_2} + f^{\dagger}_{p, E_1^+}f_{p, E_1^+} + f^{\dagger}_{p, E_1^-}f_{p,E_1^-} \nn
    &+f^{\dagger}_{p, E_2^+}f_{p, E_2^+} + f^{\dagger}_{p, E_2^-}f_{p, E_2^-}\Big)\nonumber\\
    &+ \chi\sum\limits_s\Big(\psi^{\dagger}_{s, E^+}f_{s, E^+} +\psi^{\dagger}_{s, E^-}f_{s, E^-}+ h.c.\Big)\Bigg]\nonumber\\
    + \frac{t}{2}&\sum_s(f^{\dagger}_sf_s+\psi^{\dagger}_s\psi_s),
\end{align}
where $\chi$ is a parameters that gives the Higgs hybridization. Intuitively, this hybridization is necessary because the unhybridized mean-field state has no $\psi$ parton and five $f$ partons per unit cell, and a naive projection of this unhybridized state into the physical Hilbert space would give a trivial result. We also require $\chi$ to be relatively small so that the charge distribution is not affected by the Higgs hybridization.

\begin{figure}[h]
  \centering
    \includegraphics[width=.5\textwidth]{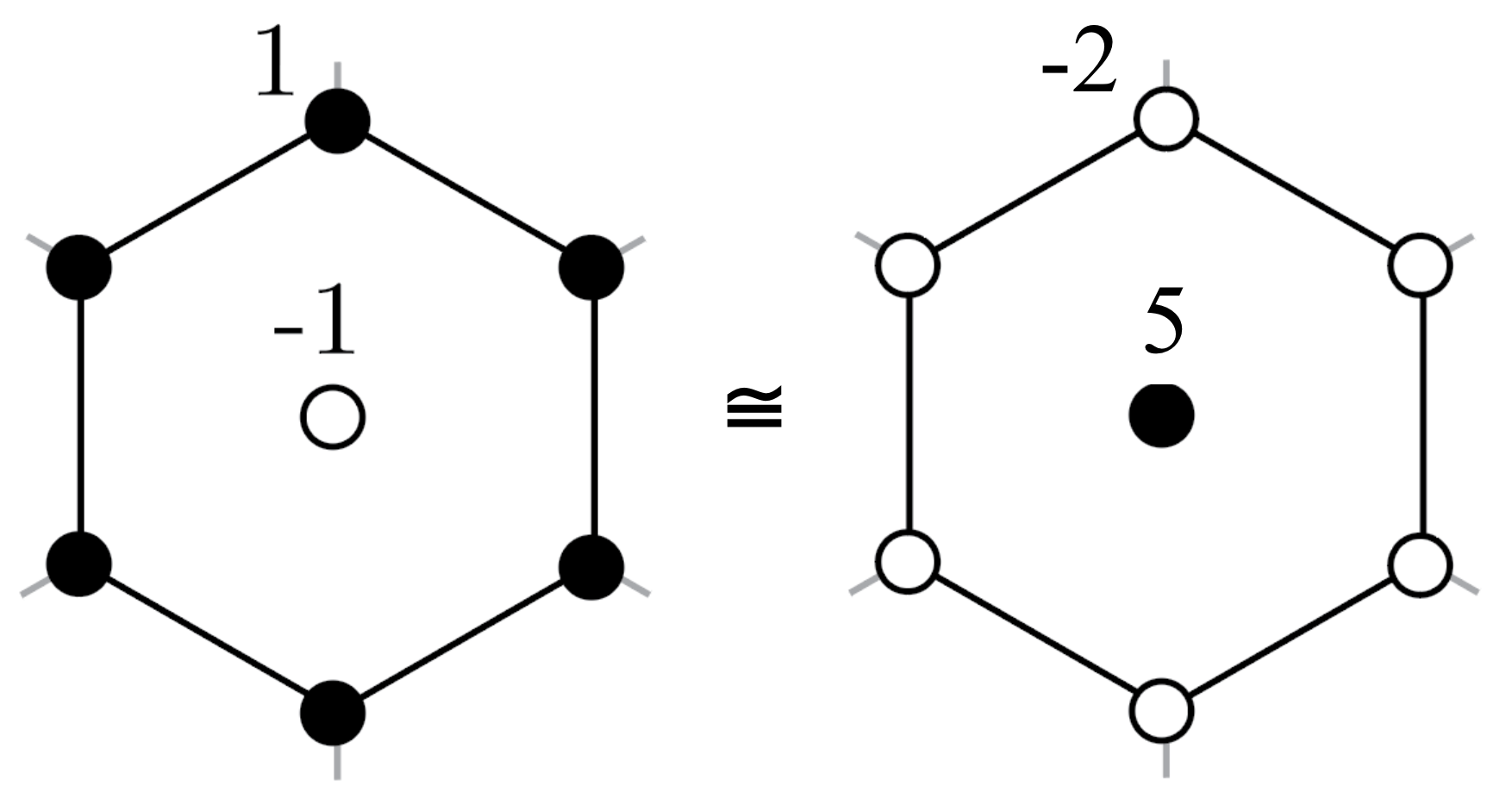}
    \caption{Right: the charge distribution from the mean field Hamiltonian Eq.~\eqref{hc4MF}, using the parton decomposition $c_s=f_{s,A_1}f_{s,E^+}f_{s,E^-}\psi^{\dagger}_{s,E^+}\psi^{\dagger}_{s,E^-}$. Left: the charge distribution from the same theory, but with partons redefined (or relabeled) using charge conjugation: $c_s = f^{\dagger}_{s, A_1}f^{\dagger}_{s, E^+}f^{\dagger}_{s, E^-}\psi_{s, E^+}\psi_{s, E^-}$. The two should be equivalent since we have merely relabeled the partons. This is indeed true since the two charge distributions are connected through lattice homotopy\cite{LatticeHomotopy}.}
  \label{hclatthomo}
\end{figure}

One can make the following observation: if we redefine the partons in the decomposition Eq.~\eqref{Parton3} as their charge conjugation, namely $c_s = f^{\dagger}_{s, A_1}f^{\dagger}_{s, E^+}f^{\dagger}_{s, E^-}\psi_{s, E^+}\psi_{s, E^-}$, then the mean field Hamiltonian will flip sign. Now the ground state has one $f$ particle occupied at each plaquette (the $f_{p,A_1}$ orbit), and two $\psi$ fermions occupied at each site, followed by appropriate $f$-$\psi$ hybridization. The parton decomposition requires that $f$ and $\psi$ fermions carry $-1$ charge (instead of $+1$ previously), and the parton vacuum (the state with $f$ and $\psi$ unoccupied) should carry physical charge $+3$. This means that the plaquette center should have charge $-1$ and the site should have charge $3-2=1$ (Fig.~\ref{hclatthomo} left). This appears to be a different charge distribution than we obtained before -- but it should not because we have merely relabeled the partons. The key is to realize that the charge at a $C_n$ rotation center is sharply defined only modulo $n$, since one can always group $n$ charges symmetrically distributed away from the center and count them in. In this particular case, the two seemingly different charge distributions differ by taking a group of $+6$ charges at each plaquette and distribute them symmetrically to the sites (Fig.~\ref{hclatthomo}). The bosonic version of this state has been discussed in Ref.~\cite{ElsePoWatanabe}.

\subsection{Square lattice}

\begin{figure}[h]
  \centering
    \includegraphics[width=.5\textwidth]{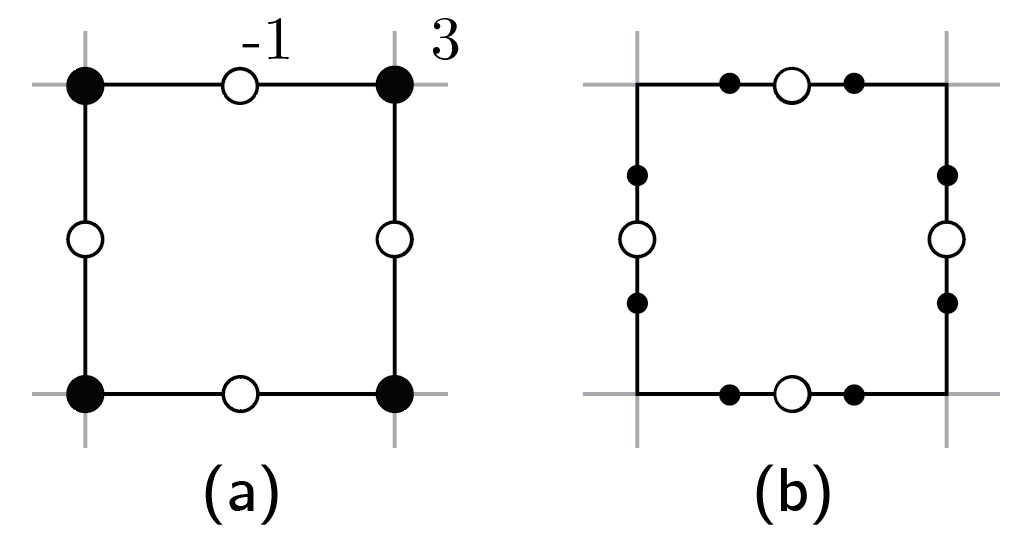}
    \caption{(a) Physical $U(1)$ charge distribution for parton construction on the square lattice. (b) Distributing parton orbitals on lattice sites to orbitals centered around the links. Each link has two nearest lattice site neighbors.}
  \label{sqlatt}
\end{figure}

There is no non-interacting, gapped model of spinless fermions in s-orbitals centered on the links of a 2D square lattice with a filling of one fermion per unit cell \cite{Bradlyn2017}. In order to create a strongly correlated state with these properties, we will again make use of the hybridized orbital technique, beginning with the charge pattern shown in Fig.~\ref{sqlatt}(a). The four orbitals derived from the partons on the sites decompose into the following irreps of lattice site symmetry group $C_{4v}$:
\begin{equation}
    \label{sqorbs}
    A_1\oplus B_1\oplus E
\end{equation}
In order to create a physical fermion operator, we allow an additional (initially empty) orbital to sit on each link in a $B_1$ irrep, which we will label as $B_1'$. The parton construction is:
\begin{equation}
    \label{sqparton}
    c_{\ell} = f_{\ell, A_1}f_{\ell, B_1}f^{\dagger}_{\ell, B_1'},
\end{equation}
where $\ell$ labels a link and the irreps $A_1$ and $B_1$ are for the link point group $C_{2v}$. This time, the $f_{\ell, B_1'}$ and $f_{\ell, B_1}$ parton orbitals must be hybridized:
\begin{align}
    H_{MF} = -t\Bigg[&\sum\limits_s\Big(f^{\dagger}_{s,A_1}f_{s,A_1} + f^{\dagger}_{s,E^+}f_{s,E^+} + f^{\dagger}_{s,E^-}f_{s,E^-}\Big)\nonumber\\
    + \chi&\sum\limits_{\ell}\Big(f^{\dagger}_{\ell,B_1'}f_{\ell,B_1} + h.c.\Big)\Bigg] + \frac{t}{2}\sum_sf^{\dagger}_sf_s.
\end{align}

\subsection{A $3D$ example}
\label{3Dexample}

Conceptually there is no obstacle to generalize our parton approach to $3D$ systems. Below we give a simple example in $3D$.

\begin{figure}[h]
  \centering
    \includegraphics[width=.5\textwidth]{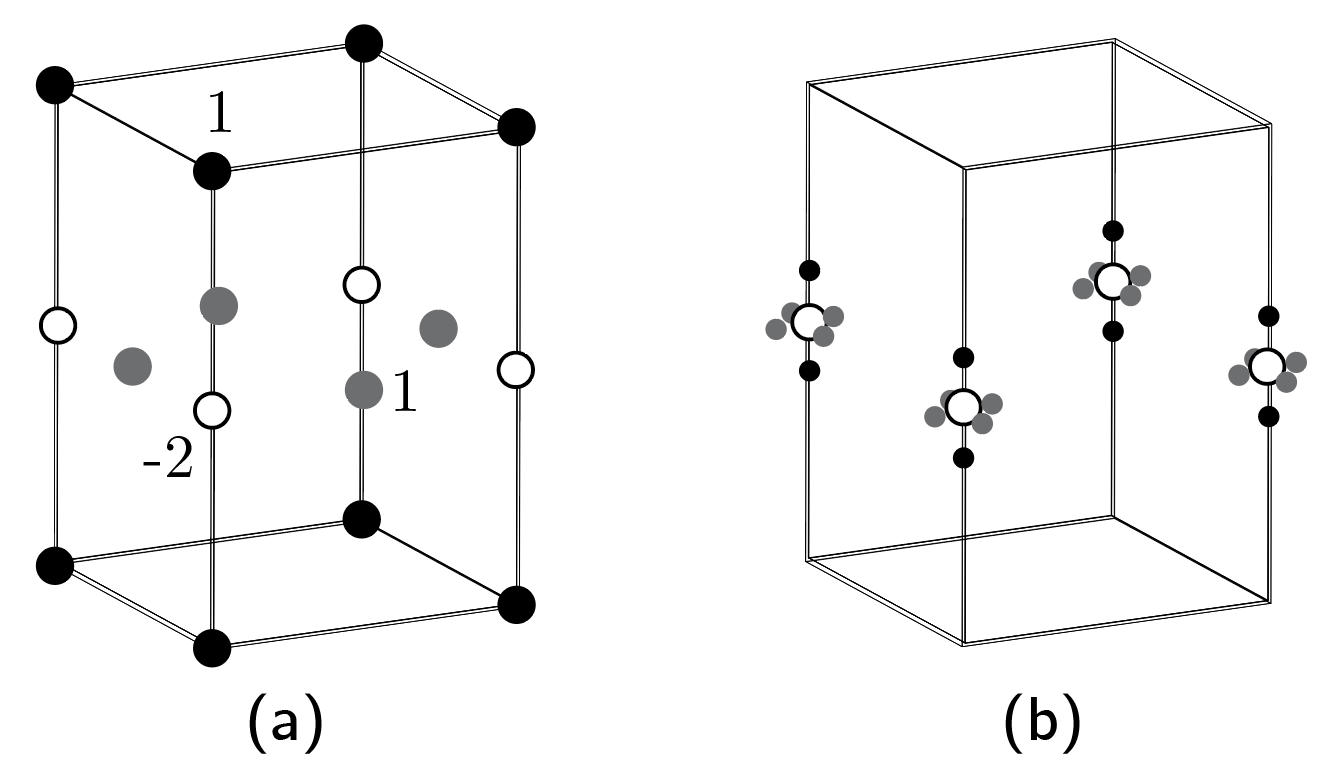}
    \caption{(a) Physical $U(1)$ charge distribution for parton construction on the tetragonal lattice. (b) Distributing parton orbitals on faces to orbitals centered around the links. Each link has two nearest vertex neighbors (black orbitals) and four nearest lateral face neighbors (grey orbitals). The unfilled orbital on the link in this figure is schematic; as detailed in the main text, we require four on-site orbitals for this construction.}
  \label{tetrlatt}
\end{figure}

There is no non-interacting, gapped model of spinless fermions in orbitals transforming like the $E_g$ irrep (a two-dimensional representation which is even under spatial inversion, i.e. transforming like a pseudovector basis) of point group $D_{4h}$ on the vertical links of a 3D tetragonal lattice with a filling of one fermion per unit cell \cite{Bradlyn2017}. We begin with one $f$ parton at each of the four lateral faces of the unit cell and one at each of the eight vertices, and ``smear" them in the manner shown in Fig.~\ref{tetrlatt}(b): the partons from the faces are distributed equally between the two nearest links, and likewise for those from the vertices. This gives a total of six orbitals associated with each link, and these decompose into irreps of the site symmetry group $D_{4h}$ as:
\begin{equation}
    2A_{1g} \oplus B_{1g} \oplus A_{2u} \oplus E_u
\end{equation}
Let the two copies of the $A_{1g}$ irrep 
in this decomposition be denoted $A_{1g}$ and $A_{1g}'$. We next place four additional one-dimensional orbitals on-site, two transforming as $B_{2g}$, one as $A_{2g}$, and the last as $B_{1u}$. 

The trick in this example is to notice that certain linear combinations of the one-dimensional irreps can be paired off such that they exchange in the same manner as the truly two-dimensional irreps under certain elements of $D_{4h}$. For example, under four-fold rotations about the z-axis:
\begin{equation}
    f_{\ell,A_{1g}} + f_{\ell,B_{2g}} \leftrightarrow f_{\ell,A_{1g}} - f_{\ell,B_{2g}},
\end{equation}
where $\ell$ labels the physical site on the link. We define the following linear combinations of parton orbitals:
\begin{align}
\label{tetrlincombns}
    \frac{1}{\sqrt{2}}\Big(f_{\ell,A_{1g}} + f_{\ell,B_{2g}}\Big)&\equiv f_{\ell, \alpha_1}\nonumber\\
    \frac{1}{\sqrt{2}}\Big(f_{\ell,A_{1g}} - f_{\ell,B_{2g}}\Big)&\equiv f_{\ell, \beta_1}\nonumber\\
    \frac{1}{\sqrt{2}}\Big(f_{\ell,A_{2g}} + f_{\ell,B_{1g}}\Big)&\equiv f_{\ell, \alpha_2}\nonumber\\
    \frac{1}{\sqrt{2}}\Big(f_{\ell,A_{2g}} - f_{\ell,B_{1g}}\Big)&\equiv f_{\ell, \beta_2}\nonumber\\
    \frac{1}{\sqrt{2}}\Big(f_{\ell,A_{2u}} + f_{\ell,B_{1u}}\Big)&\equiv f_{\ell, \alpha_3}\nonumber\\
    \frac{1}{\sqrt{2}}\Big(f_{\ell,A_{2u}} - f_{\ell,B_{1u}}\Big)&\equiv f_{\ell, \beta_3}\nonumber\\
\end{align}
Then we propose the parton construction:
\begin{align}
    \label{tetrparton}
    c_{\ell, E_g^x} &= f_{\ell, E_u^x}f_{\ell, \alpha_1}f_{\ell, \alpha_2}f^{\dagger}_{\ell, \alpha_3}f^{\dagger}_{\ell, \alpha_1'}\nonumber\\
    c_{\ell, E_g^y} &= f_{\ell, E_u^y}f_{\ell, \beta_1}f_{\ell, \beta_2}f^{\dagger}_{\ell, \beta_3}f^{\dagger}_{\ell, \beta_1'}
\end{align}
We show in Appendix~\ref{3Dxform} that the constructions Eq.~\eqref{tetrparton} indeed transform as the $E_g$ irrep of $D_{4h}$. 


We are now ready to write down our mean-field Hamiltonian:
\begin{align}
    H_{MF} = -t\Bigg[&\sum\limits_f f^{\dagger}_{f, A_g}f_{f, A_g} + \sum\limits_v f^{\dagger}_{v, A_{1g}}f_{v, A_{1g}}\nonumber\\
    + \chi&\sum\limits_{\ell}\Big(f^{\dagger}_{\ell, \alpha_1}f_{\ell, \alpha_1'} + f^{\dagger}_{\ell, \beta_1}f_{\ell, \beta_1'} + h.c.\Big)\Bigg]\nonumber\\
    + \frac{t}{2}&\sum_{\ell}f^{\dagger}_{\ell}f_{\ell},
\end{align}
where $f$ is the summation index for lateral faces (site symmetry group $D_{2h}$) and $v$ is the summation index for vertices (site symmetry group $D_{4h}$). The $\chi$ term is a symmetry-allowed hybridazation that ensures the triviality of IGG, as can be checked in a way similar to Appendix.~\ref{IGG}.

\section{Spin-$1/2$ electrons}
\label{spinhalf}

So far we have focused on FTIs of spinless particles. Motivated by real-life electrons, we now move on to discuss spin-$1/2$ fermions, assuming full $SU(2)$ spin rotation symmetry in addition to $U(1)$ charge conservation and lattice symmetries. Similar to the spinless case, an atomic or fragile phase is labeled by the symmetry quantum numbers at each Wyckoff position. The only totally symmetric representation of the group $SU(2)$ is the spin-$0$ singlet, which in turn requires the electric charge to be an even integer, since an odd integer charge necessarily carries half-integer spin. Therefore an atomic or fragile insulator of spin-$1/2$ fermions is always equivalent to that of spinless, charge-$2$ bosons (Cooper pairs). The discussions from Sec.~\ref{honeycomb} and \ref{examples} applies straightforwardly to this case: the only modification is to replace the spinless fermion operator $c$ by the Cooper pair operator $b=c_{\uparrow}c_{\downarrow}$ and change the partons $f$ from fermionic to bosonic. For example, one can construct a state represented again by Fig.~\ref{FTIillu}, but with charge $+2$ one each edge and $-2$ on each vertex.

Another common way to construct a strongly correlated insulator, motivated by realistic systems with strong local correlations, is to have exactly one electron localized at each lattice site, which effectively converts the electron system to a spin system -- this is sometimes called a ``strong Mott insulator". A fully symmetric strong Mott insulator would corresponds to a quantum paramagnet of the spin system. If this paramagnetic insulator is short-range entangled and has no protected edge mode, then from our discussion above it should be adiabatically equivalent to a Cooper pair insulator, either atomic or fragile\footnote{Notice that there are plenty of counter-examples if either of the two conditions is violated: a gapless spin liquid (long-range entangled) or a spin quantum Hall state (with protected gapless edge state) cannot be adiabatically connected to a Cooper pair insulator, since the latter must necessarily contain only gapped spin excitations, both in the bulk and on the edge.}. A paramagnetic insulator adiabatically connected to a fragile topological insulator of Cooper pairs must necessarily possess some nontrivial many-body entanglement. We therefore dub such paramagnetic insulators as ``fragile topological paramagnets".

Now several questions naturally arise: how do we decide what type of Cooper pair insulator is equivalent to a given paramagnet? Can we view all Cooper pair insulators (atomic or fragile) as some kind of quantum paramagnet? If not, what is the criterion? We answer these questions below. We will focus on the case of two spatial dimensions, although some of the arguments can be straightforwardly generalized to higher dimensions.

\subsection{Fragile topological paramagnets}

If an atomic or fragile insulator can be adiabatically deformed to a quantum paramagnet, say with one electron per lattice site, then this insulator must have the same topological response properties as the quantum paramagnet. In fact, since an SPT order is expected to be fully characterized by its topological response under various symmetry defects\cite{KapustinCobordism,FermionCobordism,ThorngrenElse}, an atomic or fragile insulator is adiabatcially connected to a quantum paramagnet if the two have the same topological response. 

An important type of topological response is probed by inserting a $2\pi$ flux into the system, and asking how the symmetry eigenvalues (quantum numbers) of the entire system change -- in other words, we measure the symmetry quantum numbers of a monopole insertion operator. The monopole quantum numbers are determined by the topological properties of the underlying insulator. The calculation is particularly simple when the insulator is atomic or fragile\cite{Songetal,shiftinsulator}: the rotation symmetry eigenvalue $\theta$ (i.e. lattice angular momentum) of the monopole with respect to a $C_n$ rotation center $r$ is given by
\be
\theta(C_n^r)=e^{iq_r2\pi/n},
\ee
where $q_r$ is the $U(1)$ charge effectively localized at the center $r$. Physically this simply comes from the Aharonov-Bohm phase as the monopole (or flux) moves around the charge. The lattice momentum is then determined by combining rotations around different centers. The monopole transforms trivially under the other (internal) symmetries like time-reversal or spin rotation, since these symmetries do not participate in the SPT order for an atomic or fragile insulator.

A quantum paramagnet with one electron per site by definition has $q_r=1$ on the lattice sites and $q_r=0$ at all the other Wyckoff positions. Therefore if an atomic or fragile insulator can be deformed to a quantum paramagnet, we must have
\bea
\label{ParaCriterion}
q_r&=&1 \hspace{5pt} ({\rm{mod }}\ n) \hspace{10pt} {\rm{on\ lattice\ sites}}, \nn
q_r&=&0  \hspace{5pt} ({\rm{mod }}\ n) \hspace{10pt} {\rm{at\ all\ other\ rotation\ centers}}, \nn
q_r&=&0  \hspace{5pt} ({\rm{mod }}\ 2) \hspace{10pt} {\rm{everywhere}}.
\eea
Eq.~\eqref{ParaCriterion} is a necessary condition for an atomic or fragile insulator to be adiabatically deformable to a quantum paramagnet. We also expect it to be sufficient, since there is no other obstruction from the topological response perspective. The condition Eq.~\eqref{ParaCriterion} can also be obtained from the ``lattice homotopy" approach\cite{ElsePoWatanabe}. Notice that the first and third line in Eq.~\eqref{ParaCriterion} implies that the lattice site cannot have a $C_2$ symmetry, which is consistent with generalized Lieb-Schultz-Mattis theorems\cite{LatticeHomotopy}. 

It turns out the fragile insulator picturized in Fig.~\ref{FTIillu} with charge $2$ on the edges and charge $(-2)$ at the vertices of a honeycomb lattice (the bosonic version of the example in Sec.~\ref{honeycomb}), is an example of fragile topological paramagnet, as it satisfies Eq.~\eqref{ParaCriterion}. Another example, also on a half-filled honeycomb lattice, is to have charge $6$ at each hexagon center and charge $(-2)$ at each vertex, as this charge distribution also satisfies Eq.~\eqref{ParaCriterion} with one electron per site on everage -- this is simply the bosonic version of the example in Sec.~\ref{simpleEx}. An example of FTI not adiabatically connected to a paramagnet is the bosonic version of the example in Sec.~\ref{lesssimpleEx}, with charge $-2$ on the plaquette and $+2$ on the site. One can show that any quantum paramagnet realized on a half-filled honeycomb lattice cannot be completely atomic -- it has to be at least fragile. The proof is simple: suppose we have an atomic insulator, with charge $q_v$ at each vertex, $q_e$ on each edge, and $q_h$ at each hexagon center. Half-filling means that $2q_v+3q_e+q_h=2$. With these constraints, it is easy to see that Eq.~\eqref{ParaCriterion} cannot be satisfied with only positive charges.

The criteria in Eq.~\eqref{ParaCriterion} may appear to imply that all topological paramagnets in a given system are equivalent, since they all have the same charge response. However this is not true since they may have different spin response. Specifically, one can probe the system by inserting a $2\pi$ flux seen only by the spin-up fermions $f_{\uparrow}$. This can be viewed as a combination of $\pi$ flux for the charge $U(1)$ symmetry and $\pi$ flux for the $S_z$ spin $U(1)$ symmetry\footnote{Or more systematically, a $\pi$ flux for the charge $U(1)$ symmetry, combined with a $\mathbb{Z}_2$ flux for the spin $SO(3)$ symmetry in which the second Stiefel-Whitney class $w_2^{SO(3)}$ is nontrivial.}. Now consider a $C_2$ rotation around a center $r$ (which cannot be the lattice site from Eq.~\eqref{ParaCriterion}) on this spin-charge flux. A simple analysis gives the $C_2$ eigenvalue of the flux
\be
\label{spinchargetheta}
\theta_{spin-charge}(C_2^r)=e^{iq_r\pi/2},
\ee
which can differentiate $q_r=0$ from $q_r=2$ mod $4$. For example, the bosonic version of the example in Sec.~\ref{honeycomb} has $q_e=2, q_h=0$ while the bosonic version of the example in Sec.~\ref{simpleEx} has $q_e=0, q_h=6$. By Eq.~\eqref{spinchargetheta} these two are clearly distinct fragile topological paramagnets. The simplest example wavefunction constructed in Ref.~\cite{RanFeatureless} corresponds to the former, with $q_v=-2, q_e=2, q_h=0$.

\section{Summary}

In this work we have developed a parton approach to construct and study various types of fragile topological insulators (FTIs), including those that cannot be realized within free fermion band theory. The basic idea is that partons can effectively carry negative charges, even when all the physical particles are positively charged. Those negatively charged partons then serve as the source of the negative charges localized at certain Wyckoff positions in real space. The parton gauge symmetry in our construction is completely Higgsed by design, so that the resulting many-body state is short-range entangled, as required for an FTI. 

Our construction appears to be quite general, and leads us to conjecture (without proof) that an FTI that cannot be realized within band theory can always be realized in interacting systems, as long as the underlying lattice Hilbert space remains nontrivial at the corresponding filling fraction. If this conjecture is correct, then the theory of fragile phases will be conceptually much simpler in interacting systems than in free fermions. In free fermion theory, to describe a possible FTI in real space, we need to specify not only the total quantum numbers at each Wyckoff position, but also the quantum numbers of each individual orbit at that position. We then have to decide whether that real space configuration can actually be realized in band theory. These make the systematic theory of FTI in band theory rather complicated (see for example Ref.~\cite{SongMonoids}). In contrast, for interacting systems an FTI is specified only by the total quantum numbers at each Wyckoff position\cite{shiftinsulator}. Furthermore once the quantum numbers are specified, we expect the state to be realizable in some interacting systems since it will not be difficult, in principle, to engineer a parton construction similar to the ones discussed in this work. At a more general level, this fits into the (again unproven) expectation that as long as a state is not explicitly forbidden by general arguments (such as Lieb-Schultz-Mattis-like theorems\cite{LSM,Hasting,LatticeHomotopy}), it should be realizable in some sufficiently generic local Hamiltonian systems. 

We should note that parton constructions in general do not lead to simple lattice Hamiltonians that can be readily solved to realize the corresponding phases. This is also true in our study, which only serves as a proof of principle that those FTIs we constructed exist for \textit{some} local Hamiltonians. Our parton constructions do motivate some variational Gutzwiller-projected wavefunctions that can be numerically tested. In fact the example in Sec.~\ref{honeycomb} has been shown in Ref.~\cite{RanFeatureless} to be a symmetric short-range entangled state.

We have also discussed FTIs in spin-$1/2$ electron systems, which can always be adiabatically viewed as FTIs made of charge-$2$, spin singlet, bosonic Cooper pairs. Amusingly, certain quantum paramagnets discussed previously in the literature (also known as ``featureless Mott insulators")\cite{RanFeatureless} are adiabatically connected to such FTIs -- we therefore dub these states ``fragile topological paramagnets".

{\textbf{Acknowledgements}}: We thank Liujun Zou for helpful discussions. Research at Perimeter Institute is supported by the Government of Canada through the Department of Innovation, Science and Economic Development Canada and by the Province of Ontario through the Ministry of Research, Innovation and Science.

\appendix

\section{Symmetry indicator analysis}
\label{Sindicator}

We show that the FTI in Sec.~\ref{honeycomb} (Fig.~\ref{FTIillu}) cannot be realized within free fermion band theory, irrespective of the details of the underlying microscopic Hilbert space (as long as it is compatible with the symmetry and charge positivity). 

We use the symmetry indicator technique\cite{indicators,Bradlyn2017}. Essentially we ask the question: if the FTI is realized using some free fermion band structure, what would the lattice symmetry eigenvalues look like in the Brillouin zone? We shall see that the charge distribution in Fig.~\ref{FTIillu} necessarily leads to an inconsistency in the symmetry eigenvalues, therefore a band theory realization of the FTI does not exist.

It turns out that we only need to focus on the lattice $C_2$ symmetry, say with respect to an edge center. The lattice momenta that are invariant under $C_2$ are the $\Gamma$ ($(k_1,k_2)=(0,0)$) and the three $M$ ($(k_1,k_2)=(0,\pi),(\pi,0),(\pi,\pi)$) points. The $C_2$ eigenvalues of the Bloch states are well defined at these four momenta. The three $+1$ charges on the edge in each unit cell requires at least three bands to realize, and it is straightforward to work out their $C_2$ eigenvalues in momentum space: $\eta\times \{1,e^{ik_1},e^{ik_2}\}$, where $\eta=\pm1$ is an orbital angular momentum that can be assigned by hand. The two $-1$ charges on the lattice sites in each unit cell requires two bands, with $C_2$ eigenvalues $\{1,-1\}$ independent of momentum. Now if a band structure realizes the FTI, its symmetry eigenvalue spectrum should correspond to the difference between the positively and negatively charged bands\cite{Pofragile}. But it is obvious that the two bands cannot be reasonably ``subtracted": for example, at $\Gamma$, their difference contains ``$-1$" copies of the $-\eta$ eigenvalue, which does not make sense for a band structure.

To complete the argument, we should also analyze the more general situation, where on each edge there are $n\geq1$ positive charges and $(n-1)$ negative charges, but with different orbital angular momentum -- without loss of generality we can take all positive bands to have $\eta=1$ and all negative bands to have $\eta=-1$. Similarly on each lattice sites there are $l\geq1$ negative charges and $(l-1)$ positve charges, and on each hexagon plaquette center $m\geq0$ positive charges (say with orbital $C_2$ angular momentum $\eta'$) and $m$ negative charges (with $-\eta'$). Repeating the above analysis, we find that the number of $+1$ and $-1$ eigenvalues at $\Gamma$ is
\bea
N^{\Gamma}_+&=&3n+\eta' m-1, \nn
N^{\Gamma}_-&=&-3n-\eta' m+2,
\eea
which could be all non-negative if $\eta'=-1$, $n=1$ and $m\in\{1,2\}$. However, at $(k_1,k_2)=(\pi,\pi)$ we have
\bea
N^{M}_+&=&-n+\eta' m+1, \nn
N^{M}_-&=&n-\eta' m,
\eea
so if we take the above solution for $\eta',n,m$ we will necessarily have $N^{M}_+<0$. Therefore a sensible band structure that gives the FTI in Fig.~\ref{FTIillu} does not exist.

\section{Gauge group analysis}
\label{IGG}

We verify that the mean field ansatz discussed in Sec.~\ref{honeycomb} has a trivial IGG. We can rewrite the mean-field Hamiltonian Eq.~\eqref{hmf} as:
\begin{equation}
    H_{MF}=\sum\limits_{ij}\sum\limits_{ab}(T_{ij})_{ab}\Psi^{\dagger}_{a,i}\Psi_{b,j},
\end{equation}
where
\begin{equation}
    \Psi_i = \Big(f_{1,i},f_{2,i},f_{3,i},f^{\dagger}_{1,i},f^{\dagger}_{2,i},f^{\dagger}_{3,i}\Big)^T
\end{equation}
and
\begin{align}
    \label{ansatz'}
    T_{ij}=\begin{pmatrix}
        -t/2&0&0&0&0&0\\
        0&0&0&0&0&0\\
        0&0&0&0&0&0\\
        0&0&0&t/2&0&0\\
        0&0&0&0&0&0\\
        0&0&0&0&0&0\\
    \end{pmatrix}&\quad ij=\includegraphics[scale=0.3, trim=0 1cm 0 0]{link1.png}\nonumber\\
    \begin{pmatrix}
        0&0&0&0&0&0\\
        0&-t/2&0&0&0&0\\
        0&0&0&0&0&0\\
        0&0&0&0&0&0\\
        0&0&0&0&t/2&0\\
        0&0&0&0&0&0\\
    \end{pmatrix}&\quad ij=\includegraphics[scale=0.3, trim=0 0.8cm 0 0]{link2.png}\nonumber\\
    \begin{pmatrix}
        0&0&0&0&0&0\\
        0&0&0&0&0&0\\
        0&0&-t/2&0&0&0\\
        0&0&0&0&0&0\\
        0&0&0&0&0&0\\
        0&0&0&0&0&t/2\\
    \end{pmatrix}&\quad ij=\includegraphics[scale=0.3, trim=0 0.8cm 0 0]{link3.png}.\nonumber\\
\end{align}
We want to work in the basis where the SU(3) gauge symmetry Eq.~\eqref{newsu3} is manifest, which will be denoted by primes:
\begin{equation}
    \Psi'_i = \Big(f^{\dagger}_{0,i},f_{+,i},f_{-,i},f_{0,i},f^{\dagger}_{+,i},f^{\dagger}_{-,i}\Big)^T
\end{equation}
Under this basis transformation the Hamiltonian looks like:
\begin{equation}
    H_{MF}=\sum\limits_{ij}\sum\limits_{ab}(T'_{ij})_{ab}\Psi'^{\dagger}_{a,i}\Psi'_{b,j},
\end{equation}
where
\begin{align}
    \label{ansatzappendix}
    T'_{ij}=-\frac{t}{6}\times\begin{pmatrix}
        -1&0&0&0&-1&-1\\
        -1&0&0&0&-1&-1\\
        -1&0&0&0&-1&-1\\
        0&1&1&1&0&0\\
        0&1&1&1&0&0\\
        0&1&1&1&0&0\\
    \end{pmatrix}&\quad ij=\includegraphics[scale=0.3, trim=0 1cm 0 0]{link1.png}\nonumber\\
    \begin{pmatrix}
        -1&0&0&0&-\overline{\omega}&-\omega\\
        -1&0&0&0&-\overline{\omega}&-\omega\\
        -1&0&0&0&-\overline{\omega}&-\omega\\
        0&\overline{\omega}&\omega&1&0&0\\
        0&\overline{\omega}&\omega&1&0&0\\
        0&\overline{\omega}&\omega&1&0&0\\
    \end{pmatrix}&\quad ij=\includegraphics[scale=0.3, trim=0 0.8cm 0 0]{link2.png}\nonumber\\
    \begin{pmatrix}
        -1&0&0&0&-\omega&-\overline{\omega}\\
        -1&0&0&0&-\omega&-\overline{\omega}\\
        -1&0&0&0&-\omega&-\overline{\omega}\\
        0&1&\overline{\omega}&\omega&0&0\\
        0&1&\overline{\omega}&\omega&0&0\\
        0&1&\overline{\omega}&\omega&0&0\\
    \end{pmatrix}&\quad ij=\includegraphics[scale=0.3, trim=0 0.8cm 0 0]{link3.png}.\nonumber\\
\end{align}
The only SU(3) element $U_i$ satisfying:
\begin{equation}
    \lp\begin{array}{c c} U_i^{\dagger}, U_i \end{array}\rp T'_{ij}\lp\begin{array}{c}
   U_j \\
   U_j^{\dagger} \end{array}\rp
\end{equation}
is the identity (this holds straightforwardly for the all terms in $H_{MF}$ associated with different link orientations). The IGG is therefore trivial, as claimed in the main text.

\section{Parton transformations for the 3D example}
\label{3Dxform}
In Table~\ref{3Dxformtable}, we list the representations of group elements for the orbitals defined in \eqref{tetrlincombns}, as well as for the two-dimensional representations of the point group $D_{4h}$. 

\begin{table}[ht]
\label{3Dxformtable}
\begin{tabular}{c c c c c c c c c c c}
\toprule[1.5pt]
& $E$ & $C_4$ & $C_2$ & $C_2'$ & $C_2''$ & $i$ & $S_4$ & $\sigma_h$ & $\sigma_v$ & $\sigma_d$ \\
\hline
($\alpha_1$,$\beta_1$)&$\mathbbm{1}$&$\tau_x$&$\mathbbm{1}$&$\tau_x$&$\mathbbm{1}$&$\mathbbm{1}$&$\tau_x$&$\mathbbm{1}$&$\tau_x$&$\mathbbm{1}$ \\
($\alpha_2$,$\beta_2$)&$\mathbbm{1}$&$\tau_x$&$\mathbbm{1}$&$-\tau_x$&$-\mathbbm{1}$&$\mathbbm{1}$&$\tau_x$&$\mathbbm{1}$&$-\tau_x$&$-\mathbbm{1}$ \\
($\alpha_3$,$\beta_3$)&$\mathbbm{1}$&$\tau_x$&$\mathbbm{1}$&$-\tau_x$&$-\mathbbm{1}$&$-\mathbbm{1}$&$-\tau_x$&$-\mathbbm{1}$&$\tau_x$&$\mathbbm{1}$ \\
($E_u^x$,$E_u^y$)&$\mathbbm{1}$&$-i\tau_y$&$\mathbbm{1}$&$-\tau_x$&$\tau_z$&$-\mathbbm{1}$&$i\tau_y$&$\mathbbm{1}$&$\tau_x$&$\tau_z$ \\
($E_g^x$,$E_g^y$)&$\mathbbm{1}$&$-i\tau_y$&$\mathbbm{1}$&$-\tau_x$&$\tau_z$&$\mathbbm{1}$&$-i\tau_y$&$-\mathbbm{1}$&$-\tau_x$&$-\tau_z$ \\
\bottomrule[1.5pt]
\end{tabular}
\caption{Transformations of orbital pairs, listed in leftmost column, under elements of the group $D_{4h}$, listed in topmost row. The symbol $\mathbbm{1}$ denotes the 2x2 identity matrix, and $(\tau_x, \tau_y, \tau_z)$ are the standard Pauli matrices. The $\alpha$ and $\beta$ orbitals are defined in Eq.~\eqref{tetrlincombns}.}
\end{table}
Using these data, we are able to work out the manner in which the physical fermion operators \eqref{tetrparton} transform. For example, under four-fold rotations about the z-axis:
\begin{align}
    f_{\ell, E_u^x}f_{\ell, \alpha_1}f_{\ell, \alpha_2}f^{\dagger}_{\ell, \alpha_3}f^{\dagger}_{\ell, \alpha_1'} &\rightarrow f_{\ell, E_u^y}f_{\ell, \beta_1}f_{\ell, \beta_2}f^{\dagger}_{\ell, \beta_3}f^{\dagger}_{\ell, \beta_1'} \nonumber\\
    f_{\ell, E_u^y}f_{\ell, \beta_1}f_{\ell, \beta_2}f^{\dagger}_{\ell, \beta_3}f^{\dagger}_{\ell, \beta_1'} &\rightarrow -f_{\ell, E_u^x}f_{\ell, \alpha_1}f_{\ell, \alpha_2}f^{\dagger}_{\ell, \alpha_3}f^{\dagger}_{\ell, \alpha_1'}
\end{align}
This is precisely what we expect from the last row of Table~\ref{3Dxformtable}; the case for the other group elements is completely analogous.

\bibliography{Fragile}

\begin{thebibliography}{24}%
\makeatletter
\providecommand \@ifxundefined [1]{%
 \@ifx{#1\undefined}
}%
\providecommand \@ifnum [1]{%
 \ifnum #1\expandafter \@firstoftwo
 \else \expandafter \@secondoftwo
 \fi
}%
\providecommand \@ifx [1]{%
 \ifx #1\expandafter \@firstoftwo
 \else \expandafter \@secondoftwo
 \fi
}%
\providecommand \natexlab [1]{#1}%
\providecommand \enquote  [1]{``#1''}%
\providecommand \bibnamefont  [1]{#1}%
\providecommand \bibfnamefont [1]{#1}%
\providecommand \citenamefont [1]{#1}%
\providecommand \href@noop [0]{\@secondoftwo}%
\providecommand \href [0]{\begingroup \@sanitize@url \@href}%
\providecommand \@href[1]{\@@startlink{#1}\@@href}%
\providecommand \@@href[1]{\endgroup#1\@@endlink}%
\providecommand \@sanitize@url [0]{\catcode `\\12\catcode `\$12\catcode
  `\&12\catcode `\#12\catcode `\^12\catcode `\_12\catcode `\%12\relax}%
\providecommand \@@startlink[1]{}%
\providecommand \@@endlink[0]{}%
\providecommand \url  [0]{\begingroup\@sanitize@url \@url }%
\providecommand \@url [1]{\endgroup\@href {#1}{\urlprefix }}%
\providecommand \urlprefix  [0]{URL }%
\providecommand \Eprint [0]{\href }%
\providecommand \doibase [0]{http://dx.doi.org/}%
\providecommand \selectlanguage [0]{\@gobble}%
\providecommand \bibinfo  [0]{\@secondoftwo}%
\providecommand \bibfield  [0]{\@secondoftwo}%
\providecommand \translation [1]{[#1]}%
\providecommand \BibitemOpen [0]{}%
\providecommand \bibitemStop [0]{}%
\providecommand \bibitemNoStop [0]{.\EOS\space}%
\providecommand \EOS [0]{\spacefactor3000\relax}%
\providecommand \BibitemShut  [1]{\csname bibitem#1\endcsname}%
\let\auto@bib@innerbib\@empty
\bibitem [{\citenamefont {{Chen}}\ \emph {et~al.}(2013)\citenamefont {{Chen}},
  \citenamefont {{Gu}}, \citenamefont {{Liu}},\ and\ \citenamefont
  {{Wen}}}]{ChenCohomology}%
  \BibitemOpen
  \bibfield  {author} {\bibinfo {author} {\bibfnamefont {Xie}\ \bibnamefont
  {{Chen}}}, \bibinfo {author} {\bibfnamefont {Zheng-Cheng}\ \bibnamefont
  {{Gu}}}, \bibinfo {author} {\bibfnamefont {Zheng-Xin}\ \bibnamefont {{Liu}}},
  \ and\ \bibinfo {author} {\bibfnamefont {Xiao-Gang}\ \bibnamefont {{Wen}}},\
  }\bibfield  {title} {\enquote {\bibinfo {title} {{Symmetry protected
  topological orders and the group cohomology of their symmetry group}},}\
  }\href {\doibase 10.1103/PhysRevB.87.155114} {\bibfield  {journal} {\bibinfo
  {journal} {\prb}\ }\textbf {\bibinfo {volume} {87}},\ \bibinfo {eid} {155114}
  (\bibinfo {year} {2013})},\ \Eprint {http://arxiv.org/abs/1106.4772}
  {arXiv:1106.4772 [cond-mat.str-el]} \BibitemShut {NoStop}%
\bibitem [{\citenamefont {Po}\ \emph {et~al.}(2018{\natexlab{a}})\citenamefont
  {Po}, \citenamefont {Watanabe},\ and\ \citenamefont
  {Vishwanath}}]{Pofragile}%
  \BibitemOpen
  \bibfield  {author} {\bibinfo {author} {\bibfnamefont {Hoi~Chun}\
  \bibnamefont {Po}}, \bibinfo {author} {\bibfnamefont {Haruki}\ \bibnamefont
  {Watanabe}}, \ and\ \bibinfo {author} {\bibfnamefont {Ashvin}\ \bibnamefont
  {Vishwanath}},\ }\bibfield  {title} {\enquote {\bibinfo {title} {Fragile
  topology and wannier obstructions},}\ }\href {\doibase
  10.1103/PhysRevLett.121.126402} {\bibfield  {journal} {\bibinfo  {journal}
  {Phys. Rev. Lett.}\ }\textbf {\bibinfo {volume} {121}},\ \bibinfo {pages}
  {126402} (\bibinfo {year} {2018}{\natexlab{a}})}\BibitemShut {NoStop}%
\bibitem [{\citenamefont {Cano}\ \emph {et~al.}(2018)\citenamefont {Cano},
  \citenamefont {Bradlyn}, \citenamefont {Wang}, \citenamefont {Elcoro},
  \citenamefont {Vergniory}, \citenamefont {Felser}, \citenamefont {Aroyo},\
  and\ \citenamefont {Bernevig}}]{Cano_2018}%
  \BibitemOpen
  \bibfield  {author} {\bibinfo {author} {\bibfnamefont {Jennifer}\
  \bibnamefont {Cano}}, \bibinfo {author} {\bibfnamefont {Barry}\ \bibnamefont
  {Bradlyn}}, \bibinfo {author} {\bibfnamefont {Zhijun}\ \bibnamefont {Wang}},
  \bibinfo {author} {\bibfnamefont {L.}~\bibnamefont {Elcoro}}, \bibinfo
  {author} {\bibfnamefont {M.~G.}\ \bibnamefont {Vergniory}}, \bibinfo {author}
  {\bibfnamefont {C.}~\bibnamefont {Felser}}, \bibinfo {author} {\bibfnamefont
  {M.~I.}\ \bibnamefont {Aroyo}}, \ and\ \bibinfo {author} {\bibfnamefont
  {B.~Andrei}\ \bibnamefont {Bernevig}},\ }\bibfield  {title} {\enquote
  {\bibinfo {title} {Topology of disconnected elementary band
  representations},}\ }\href {\doibase 10.1103/PhysRevLett.120.266401}
  {\bibfield  {journal} {\bibinfo  {journal} {Phys. Rev. Lett.}\ }\textbf
  {\bibinfo {volume} {120}},\ \bibinfo {pages} {266401} (\bibinfo {year}
  {2018})}\BibitemShut {NoStop}%
\bibitem [{\citenamefont {Kim}\ \emph {et~al.}(2016)\citenamefont {Kim},
  \citenamefont {Lee}, \citenamefont {Jiang}, \citenamefont {Ware},
  \citenamefont {Jian}, \citenamefont {Zaletel}, \citenamefont {Han},\ and\
  \citenamefont {Ran}}]{RanFeatureless}%
  \BibitemOpen
  \bibfield  {author} {\bibinfo {author} {\bibfnamefont {Panjin}\ \bibnamefont
  {Kim}}, \bibinfo {author} {\bibfnamefont {Hyunyong}\ \bibnamefont {Lee}},
  \bibinfo {author} {\bibfnamefont {Shenghan}\ \bibnamefont {Jiang}}, \bibinfo
  {author} {\bibfnamefont {Brayden}\ \bibnamefont {Ware}}, \bibinfo {author}
  {\bibfnamefont {Chao-Ming}\ \bibnamefont {Jian}}, \bibinfo {author}
  {\bibfnamefont {Michael}\ \bibnamefont {Zaletel}}, \bibinfo {author}
  {\bibfnamefont {Jung~Hoon}\ \bibnamefont {Han}}, \ and\ \bibinfo {author}
  {\bibfnamefont {Ying}\ \bibnamefont {Ran}},\ }\bibfield  {title} {\enquote
  {\bibinfo {title} {Featureless quantum insulator on the honeycomb lattice},}\
  }\href {\doibase 10.1103/PhysRevB.94.064432} {\bibfield  {journal} {\bibinfo
  {journal} {Phys. Rev. B}\ }\textbf {\bibinfo {volume} {94}},\ \bibinfo
  {pages} {064432} (\bibinfo {year} {2016})}\BibitemShut {NoStop}%
\bibitem [{\citenamefont {Po}\ \emph {et~al.}(2018{\natexlab{b}})\citenamefont
  {Po}, \citenamefont {Zou}, \citenamefont {Vishwanath},\ and\ \citenamefont
  {Senthil}}]{TBGFragile}%
  \BibitemOpen
  \bibfield  {author} {\bibinfo {author} {\bibfnamefont {Hoi~Chun}\
  \bibnamefont {Po}}, \bibinfo {author} {\bibfnamefont {Liujun}\ \bibnamefont
  {Zou}}, \bibinfo {author} {\bibfnamefont {Ashvin}\ \bibnamefont
  {Vishwanath}}, \ and\ \bibinfo {author} {\bibfnamefont {T.}~\bibnamefont
  {Senthil}},\ }\bibfield  {title} {\enquote {\bibinfo {title} {Origin of mott
  insulating behavior and superconductivity in twisted bilayer graphene},}\
  }\href {\doibase 10.1103/PhysRevX.8.031089} {\bibfield  {journal} {\bibinfo
  {journal} {Phys. Rev. X}\ }\textbf {\bibinfo {volume} {8}},\ \bibinfo {pages}
  {031089} (\bibinfo {year} {2018}{\natexlab{b}})}\BibitemShut {NoStop}%
\bibitem [{\citenamefont {{Else}}\ \emph {et~al.}(2019)\citenamefont {{Else}},
  \citenamefont {{Po}},\ and\ \citenamefont {{Watanabe}}}]{ElsePoWatanabe}%
  \BibitemOpen
  \bibfield  {author} {\bibinfo {author} {\bibfnamefont {Dominic~V.}\
  \bibnamefont {{Else}}}, \bibinfo {author} {\bibfnamefont {Hoi~Chun}\
  \bibnamefont {{Po}}}, \ and\ \bibinfo {author} {\bibfnamefont {Haruki}\
  \bibnamefont {{Watanabe}}},\ }\bibfield  {title} {\enquote {\bibinfo {title}
  {{Fragile topological phases in interacting systems}},}\ }\href {\doibase
  10.1103/PhysRevB.99.125122} {\bibfield  {journal} {\bibinfo  {journal}
  {\prb}\ }\textbf {\bibinfo {volume} {99}},\ \bibinfo {eid} {125122} (\bibinfo
  {year} {2019})},\ \Eprint {http://arxiv.org/abs/1809.02128} {arXiv:1809.02128
  [cond-mat.str-el]} \BibitemShut {NoStop}%
\bibitem [{\citenamefont {{Liu}}\ \emph {et~al.}(2019)\citenamefont {{Liu}},
  \citenamefont {{Vishwanath}},\ and\ \citenamefont
  {{Khalaf}}}]{shiftinsulator}%
  \BibitemOpen
  \bibfield  {author} {\bibinfo {author} {\bibfnamefont {Shang}\ \bibnamefont
  {{Liu}}}, \bibinfo {author} {\bibfnamefont {Ashvin}\ \bibnamefont
  {{Vishwanath}}}, \ and\ \bibinfo {author} {\bibfnamefont {Eslam}\
  \bibnamefont {{Khalaf}}},\ }\bibfield  {title} {\enquote {\bibinfo {title}
  {{Shift Insulators: Rotation-Protected Two-Dimensional Topological
  Crystalline Insulators}},}\ }\href {\doibase 10.1103/PhysRevX.9.031003}
  {\bibfield  {journal} {\bibinfo  {journal} {Physical Review X}\ }\textbf
  {\bibinfo {volume} {9}},\ \bibinfo {eid} {031003} (\bibinfo {year} {2019})},\
  \Eprint {http://arxiv.org/abs/1809.01636} {arXiv:1809.01636
  [cond-mat.mes-hall]} \BibitemShut {NoStop}%
\bibitem [{\citenamefont {Wen}(2004)}]{wenbook}%
  \BibitemOpen
  \bibfield  {author} {\bibinfo {author} {\bibfnamefont {Xiao-Gang}\
  \bibnamefont {Wen}},\ }\href
  {http://libproxy.mit.edu/login?url=http://search.ebscohost.com/login.aspx?direct=true&db=nlebk&AN=186592&site=ehost-live}
  {\emph {\bibinfo {title} {Quantum Field Theory of Many-body Systems : From
  the Origin of Sound to an Origin of Light and Electrons.}}},\ Oxford Graduate
  Texts\ (\bibinfo  {publisher} {OUP Premium},\ \bibinfo {year}
  {2004})\BibitemShut {NoStop}%
\bibitem [{\citenamefont {Lee}\ \emph {et~al.}(2006)\citenamefont {Lee},
  \citenamefont {Nagaosa},\ and\ \citenamefont {Wen}}]{LeeNagaosaWen}%
  \BibitemOpen
  \bibfield  {author} {\bibinfo {author} {\bibfnamefont {Patrick~A.}\
  \bibnamefont {Lee}}, \bibinfo {author} {\bibfnamefont {Naoto}\ \bibnamefont
  {Nagaosa}}, \ and\ \bibinfo {author} {\bibfnamefont {Xiao-Gang}\ \bibnamefont
  {Wen}},\ }\bibfield  {title} {\enquote {\bibinfo {title} {Doping a mott
  insulator: Physics of high-temperature superconductivity},}\ }\href {\doibase
  10.1103/RevModPhys.78.17} {\bibfield  {journal} {\bibinfo  {journal} {Rev.
  Mod. Phys.}\ }\textbf {\bibinfo {volume} {78}},\ \bibinfo {pages} {17--85}
  (\bibinfo {year} {2006})}\BibitemShut {NoStop}%
\bibitem [{\citenamefont {{Kimchi}}\ \emph {et~al.}(2013)\citenamefont
  {{Kimchi}}, \citenamefont {{Parameswaran}}, \citenamefont {{Turner}},
  \citenamefont {{Wang}},\ and\ \citenamefont
  {{Vishwanath}}}]{KimchiFeatureless}%
  \BibitemOpen
  \bibfield  {author} {\bibinfo {author} {\bibfnamefont {Itamar}\ \bibnamefont
  {{Kimchi}}}, \bibinfo {author} {\bibfnamefont {S.~A.}\ \bibnamefont
  {{Parameswaran}}}, \bibinfo {author} {\bibfnamefont {Ari~M.}\ \bibnamefont
  {{Turner}}}, \bibinfo {author} {\bibfnamefont {Fa}~\bibnamefont {{Wang}}}, \
  and\ \bibinfo {author} {\bibfnamefont {Ashvin}\ \bibnamefont
  {{Vishwanath}}},\ }\bibfield  {title} {\enquote {\bibinfo {title}
  {{Featureless and nonfractionalized Mott insulators on the honeycomb lattice
  at 1/2 site filling}},}\ }\href {\doibase 10.1073/pnas.1307245110} {\bibfield
   {journal} {\bibinfo  {journal} {Proceedings of the National Academy of
  Science}\ }\textbf {\bibinfo {volume} {110}},\ \bibinfo {pages}
  {16378--16383} (\bibinfo {year} {2013})},\ \Eprint
  {http://arxiv.org/abs/1207.0498} {arXiv:1207.0498 [cond-mat.str-el]}
  \BibitemShut {NoStop}%
\bibitem [{\citenamefont {{Jian}}\ and\ \citenamefont
  {{Zaletel}}(2016)}]{JianFeatureless}%
  \BibitemOpen
  \bibfield  {author} {\bibinfo {author} {\bibfnamefont {Chao-Ming}\
  \bibnamefont {{Jian}}}\ and\ \bibinfo {author} {\bibfnamefont {Michael}\
  \bibnamefont {{Zaletel}}},\ }\bibfield  {title} {\enquote {\bibinfo {title}
  {{Existence of featureless paramagnets on the square and the honeycomb
  lattices in 2+1 dimensions}},}\ }\href {\doibase 10.1103/PhysRevB.93.035114}
  {\bibfield  {journal} {\bibinfo  {journal} {\prb}\ }\textbf {\bibinfo
  {volume} {93}},\ \bibinfo {eid} {035114} (\bibinfo {year} {2016})},\ \Eprint
  {http://arxiv.org/abs/1507.00361} {arXiv:1507.00361 [cond-mat.str-el]}
  \BibitemShut {NoStop}%
\bibitem [{\citenamefont {Song}\ \emph {et~al.}(2017)\citenamefont {Song},
  \citenamefont {Huang}, \citenamefont {Fu},\ and\ \citenamefont
  {Hermele}}]{song_2017}%
  \BibitemOpen
  \bibfield  {author} {\bibinfo {author} {\bibfnamefont {Hao}\ \bibnamefont
  {Song}}, \bibinfo {author} {\bibfnamefont {Sheng-Jie}\ \bibnamefont {Huang}},
  \bibinfo {author} {\bibfnamefont {Liang}\ \bibnamefont {Fu}}, \ and\ \bibinfo
  {author} {\bibfnamefont {Michael}\ \bibnamefont {Hermele}},\ }\bibfield
  {title} {\enquote {\bibinfo {title} {Topological phases protected by point
  group symmetry},}\ }\href {\doibase 10.1103/PhysRevX.7.011020} {\bibfield
  {journal} {\bibinfo  {journal} {Phys. Rev. X}\ }\textbf {\bibinfo {volume}
  {7}},\ \bibinfo {pages} {011020} (\bibinfo {year} {2017})}\BibitemShut
  {NoStop}%
\bibitem [{\citenamefont {Song}\ \emph {et~al.}(2020)\citenamefont {Song},
  \citenamefont {He}, \citenamefont {Vishwanath},\ and\ \citenamefont
  {Wang}}]{Songetal}%
  \BibitemOpen
  \bibfield  {author} {\bibinfo {author} {\bibfnamefont {Xue-Yang}\
  \bibnamefont {Song}}, \bibinfo {author} {\bibfnamefont {Yin-Chen}\
  \bibnamefont {He}}, \bibinfo {author} {\bibfnamefont {Ashvin}\ \bibnamefont
  {Vishwanath}}, \ and\ \bibinfo {author} {\bibfnamefont {Chong}\ \bibnamefont
  {Wang}},\ }\bibfield  {title} {\enquote {\bibinfo {title} {From spinon band
  topology to the symmetry quantum numbers of monopoles in dirac spin
  liquids},}\ }\href {\doibase 10.1103/PhysRevX.10.011033} {\bibfield
  {journal} {\bibinfo  {journal} {Phys. Rev. X}\ }\textbf {\bibinfo {volume}
  {10}},\ \bibinfo {pages} {011033} (\bibinfo {year} {2020})}\BibitemShut
  {NoStop}%
\bibitem [{\citenamefont {{Po}}\ \emph
  {et~al.}(2017{\natexlab{a}})\citenamefont {{Po}}, \citenamefont
  {{Vishwanath}},\ and\ \citenamefont {{Watanabe}}}]{indicators}%
  \BibitemOpen
  \bibfield  {author} {\bibinfo {author} {\bibfnamefont {Hoi~Chun}\
  \bibnamefont {{Po}}}, \bibinfo {author} {\bibfnamefont {Ashvin}\ \bibnamefont
  {{Vishwanath}}}, \ and\ \bibinfo {author} {\bibfnamefont {Haruki}\
  \bibnamefont {{Watanabe}}},\ }\bibfield  {title} {\enquote {\bibinfo {title}
  {{Complete theory of symmetry-based indicators of band topology}},}\ }\href
  {\doibase 10.1038/s41467-017-00133-2} {\bibfield  {journal} {\bibinfo
  {journal} {Nature Communications}\ }\textbf {\bibinfo {volume} {8}},\
  \bibinfo {eid} {50} (\bibinfo {year} {2017}{\natexlab{a}})},\ \Eprint
  {http://arxiv.org/abs/1703.00911} {arXiv:1703.00911 [cond-mat.str-el]}
  \BibitemShut {NoStop}%
\bibitem [{\citenamefont {{Bradlyn}}\ \emph {et~al.}(2017)\citenamefont
  {{Bradlyn}}, \citenamefont {{Elcoro}}, \citenamefont {{Cano}}, \citenamefont
  {{Vergniory}}, \citenamefont {{Wang}}, \citenamefont {{Felser}},
  \citenamefont {{Aroyo}},\ and\ \citenamefont {{Bernevig}}}]{Bradlyn2017}%
  \BibitemOpen
  \bibfield  {author} {\bibinfo {author} {\bibfnamefont {Barry}\ \bibnamefont
  {{Bradlyn}}}, \bibinfo {author} {\bibfnamefont {L.}~\bibnamefont {{Elcoro}}},
  \bibinfo {author} {\bibfnamefont {Jennifer}\ \bibnamefont {{Cano}}}, \bibinfo
  {author} {\bibfnamefont {M.~G.}\ \bibnamefont {{Vergniory}}}, \bibinfo
  {author} {\bibfnamefont {Zhijun}\ \bibnamefont {{Wang}}}, \bibinfo {author}
  {\bibfnamefont {C.}~\bibnamefont {{Felser}}}, \bibinfo {author}
  {\bibfnamefont {M.~I.}\ \bibnamefont {{Aroyo}}}, \ and\ \bibinfo {author}
  {\bibfnamefont {B.~Andrei}\ \bibnamefont {{Bernevig}}},\ }\bibfield  {title}
  {\enquote {\bibinfo {title} {{Topological quantum chemistry}},}\ }\href
  {\doibase 10.1038/nature23268} {\bibfield  {journal} {\bibinfo  {journal}
  {\nat}\ }\textbf {\bibinfo {volume} {547}},\ \bibinfo {pages} {298--305}
  (\bibinfo {year} {2017})},\ \Eprint {http://arxiv.org/abs/1703.02050}
  {arXiv:1703.02050 [cond-mat.mes-hall]} \BibitemShut {NoStop}%
\bibitem [{\citenamefont {Kruthoff}\ \emph {et~al.}(2017)\citenamefont
  {Kruthoff}, \citenamefont {de~Boer}, \citenamefont {van Wezel}, \citenamefont
  {Kane},\ and\ \citenamefont {Slager}}]{Slager17}%
  \BibitemOpen
  \bibfield  {author} {\bibinfo {author} {\bibfnamefont {Jorrit}\ \bibnamefont
  {Kruthoff}}, \bibinfo {author} {\bibfnamefont {Jan}\ \bibnamefont {de~Boer}},
  \bibinfo {author} {\bibfnamefont {Jasper}\ \bibnamefont {van Wezel}},
  \bibinfo {author} {\bibfnamefont {Charles~L.}\ \bibnamefont {Kane}}, \ and\
  \bibinfo {author} {\bibfnamefont {Robert-Jan}\ \bibnamefont {Slager}},\
  }\bibfield  {title} {\enquote {\bibinfo {title} {Topological classification
  of crystalline insulators through band structure combinatorics},}\ }\href
  {\doibase 10.1103/PhysRevX.7.041069} {\bibfield  {journal} {\bibinfo
  {journal} {Phys. Rev. X}\ }\textbf {\bibinfo {volume} {7}},\ \bibinfo {pages}
  {041069} (\bibinfo {year} {2017})}\BibitemShut {NoStop}%
\bibitem [{\citenamefont {{Bouhon}}\ \emph {et~al.}(2019)\citenamefont
  {{Bouhon}}, \citenamefont {{Black-Schaffer}},\ and\ \citenamefont
  {{Slager}}}]{Slager19}%
  \BibitemOpen
  \bibfield  {author} {\bibinfo {author} {\bibfnamefont {Adrien}\ \bibnamefont
  {{Bouhon}}}, \bibinfo {author} {\bibfnamefont {Annica~M.}\ \bibnamefont
  {{Black-Schaffer}}}, \ and\ \bibinfo {author} {\bibfnamefont {Robert-Jan}\
  \bibnamefont {{Slager}}},\ }\bibfield  {title} {\enquote {\bibinfo {title}
  {{Wilson loop approach to fragile topology of split elementary band
  representations and topological crystalline insulators with time-reversal
  symmetry}},}\ }\href {\doibase 10.1103/PhysRevB.100.195135} {\bibfield
  {journal} {\bibinfo  {journal} {\prb}\ }\textbf {\bibinfo {volume} {100}},\
  \bibinfo {eid} {195135} (\bibinfo {year} {2019})},\ \Eprint
  {http://arxiv.org/abs/1804.09719} {arXiv:1804.09719 [cond-mat.mes-hall]}
  \BibitemShut {NoStop}%
\bibitem [{\citenamefont {{Po}}\ \emph
  {et~al.}(2017{\natexlab{b}})\citenamefont {{Po}}, \citenamefont {{Watanabe}},
  \citenamefont {{Jian}},\ and\ \citenamefont {{Zaletel}}}]{LatticeHomotopy}%
  \BibitemOpen
  \bibfield  {author} {\bibinfo {author} {\bibfnamefont {Hoi~Chun}\
  \bibnamefont {{Po}}}, \bibinfo {author} {\bibfnamefont {Haruki}\ \bibnamefont
  {{Watanabe}}}, \bibinfo {author} {\bibfnamefont {Chao-Ming}\ \bibnamefont
  {{Jian}}}, \ and\ \bibinfo {author} {\bibfnamefont {Michael~P.}\ \bibnamefont
  {{Zaletel}}},\ }\bibfield  {title} {\enquote {\bibinfo {title} {{Lattice
  Homotopy Constraints on Phases of Quantum Magnets}},}\ }\href {\doibase
  10.1103/PhysRevLett.119.127202} {\bibfield  {journal} {\bibinfo  {journal}
  {\prl}\ }\textbf {\bibinfo {volume} {119}},\ \bibinfo {eid} {127202}
  (\bibinfo {year} {2017}{\natexlab{b}})},\ \Eprint
  {http://arxiv.org/abs/1703.06882} {arXiv:1703.06882 [cond-mat.str-el]}
  \BibitemShut {NoStop}%
\bibitem [{\citenamefont {{Kapustin}}(2014)}]{KapustinCobordism}%
  \BibitemOpen
  \bibfield  {author} {\bibinfo {author} {\bibfnamefont {Anton}\ \bibnamefont
  {{Kapustin}}},\ }\bibfield  {title} {\enquote {\bibinfo {title} {{Symmetry
  Protected Topological Phases, Anomalies, and Cobordisms: Beyond Group
  Cohomology}},}\ }\href@noop {} {\bibfield  {journal} {\bibinfo  {journal}
  {arXiv e-prints}\ ,\ \bibinfo {eid} {arXiv:1403.1467}} (\bibinfo {year}
  {2014})},\ \Eprint {http://arxiv.org/abs/1403.1467} {arXiv:1403.1467
  [cond-mat.str-el]} \BibitemShut {NoStop}%
\bibitem [{\citenamefont {{Kapustin}}\ \emph {et~al.}(2015)\citenamefont
  {{Kapustin}}, \citenamefont {{Thorngren}}, \citenamefont {{Turzillo}},\ and\
  \citenamefont {{Wang}}}]{FermionCobordism}%
  \BibitemOpen
  \bibfield  {author} {\bibinfo {author} {\bibfnamefont {Anton}\ \bibnamefont
  {{Kapustin}}}, \bibinfo {author} {\bibfnamefont {Ryan}\ \bibnamefont
  {{Thorngren}}}, \bibinfo {author} {\bibfnamefont {Alex}\ \bibnamefont
  {{Turzillo}}}, \ and\ \bibinfo {author} {\bibfnamefont {Zitao}\ \bibnamefont
  {{Wang}}},\ }\bibfield  {title} {\enquote {\bibinfo {title} {{Fermionic
  symmetry protected topological phases and cobordisms}},}\ }\href {\doibase
  10.1007/JHEP12(2015)052} {\bibfield  {journal} {\bibinfo  {journal} {Journal
  of High Energy Physics}\ }\textbf {\bibinfo {volume} {2015}},\ \bibinfo {eid}
  {52} (\bibinfo {year} {2015})},\ \Eprint {http://arxiv.org/abs/1406.7329}
  {arXiv:1406.7329 [cond-mat.str-el]} \BibitemShut {NoStop}%
\bibitem [{\citenamefont {{Thorngren}}\ and\ \citenamefont
  {{Else}}(2016)}]{ThorngrenElse}%
  \BibitemOpen
  \bibfield  {author} {\bibinfo {author} {\bibfnamefont {Ryan}\ \bibnamefont
  {{Thorngren}}}\ and\ \bibinfo {author} {\bibfnamefont {Dominic~V.}\
  \bibnamefont {{Else}}},\ }\bibfield  {title} {\enquote {\bibinfo {title}
  {{Gauging spatial symmetries and the classification of topological
  crystalline phases}},}\ }\href@noop {} {\bibfield  {journal} {\bibinfo
  {journal} {arXiv e-prints}\ ,\ \bibinfo {eid} {arXiv:1612.00846}} (\bibinfo
  {year} {2016})},\ \Eprint {http://arxiv.org/abs/1612.00846} {arXiv:1612.00846
  [cond-mat.str-el]} \BibitemShut {NoStop}%
\bibitem [{\citenamefont {{Song}}\ \emph {et~al.}(2019)\citenamefont {{Song}},
  \citenamefont {{Elcoro}}, \citenamefont {{Regnault}},\ and\ \citenamefont
  {{Bernevig}}}]{SongMonoids}%
  \BibitemOpen
  \bibfield  {author} {\bibinfo {author} {\bibfnamefont {Zhida}\ \bibnamefont
  {{Song}}}, \bibinfo {author} {\bibfnamefont {L.}~\bibnamefont {{Elcoro}}},
  \bibinfo {author} {\bibfnamefont {Nicolas}\ \bibnamefont {{Regnault}}}, \
  and\ \bibinfo {author} {\bibfnamefont {B.~Andrei}\ \bibnamefont
  {{Bernevig}}},\ }\bibfield  {title} {\enquote {\bibinfo {title} {{Fragile
  Phases As Affine Monoids: Classification and Material Examples}},}\
  }\href@noop {} {\bibfield  {journal} {\bibinfo  {journal} {arXiv e-prints}\
  ,\ \bibinfo {eid} {arXiv:1905.03262}} (\bibinfo {year} {2019})},\ \Eprint
  {http://arxiv.org/abs/1905.03262} {arXiv:1905.03262 [cond-mat.mes-hall]}
  \BibitemShut {NoStop}%
\bibitem [{\citenamefont {Lieb}\ \emph {et~al.}(1961)\citenamefont {Lieb},
  \citenamefont {Schultz},\ and\ \citenamefont {Mattis}}]{LSM}%
  \BibitemOpen
  \bibfield  {author} {\bibinfo {author} {\bibfnamefont {Elliott}\ \bibnamefont
  {Lieb}}, \bibinfo {author} {\bibfnamefont {Theodore}\ \bibnamefont
  {Schultz}}, \ and\ \bibinfo {author} {\bibfnamefont {Daniel}\ \bibnamefont
  {Mattis}},\ }\bibfield  {title} {\enquote {\bibinfo {title} {Two soluble
  models of an antiferromagnetic chain},}\ }\href {\doibase
  https://doi.org/10.1016/0003-4916(61)90115-4} {\bibfield  {journal} {\bibinfo
   {journal} {Annals of Physics}\ }\textbf {\bibinfo {volume} {16}},\ \bibinfo
  {pages} {407 -- 466} (\bibinfo {year} {1961})}\BibitemShut {NoStop}%
\bibitem [{\citenamefont {Hastings}(2004)}]{Hasting}%
  \BibitemOpen
  \bibfield  {author} {\bibinfo {author} {\bibfnamefont {M.~B.}\ \bibnamefont
  {Hastings}},\ }\bibfield  {title} {\enquote {\bibinfo {title}
  {Lieb-schultz-mattis in higher dimensions},}\ }\href {\doibase
  10.1103/PhysRevB.69.104431} {\bibfield  {journal} {\bibinfo  {journal} {Phys.
  Rev. B}\ }\textbf {\bibinfo {volume} {69}},\ \bibinfo {pages} {104431}
  (\bibinfo {year} {2004})}\BibitemShut {NoStop}%
\end{thebibliography}%

\end{document}